\documentclass[journal]{IEEEtran}

\usepackage{cite}
\usepackage{subcaption}
\usepackage{algorithm}
\usepackage[noend]{algpseudocode}
\usepackage{RobStd}
\usepackage{multirow}
\usepackage{paralist}
\usepackage{cprotect}
\usepackage{amssymb}
\usepackage{dirtytalk}
\usepackage{amsmath}
\usepackage{booktabs} % For formal tables
\usepackage{graphicx}

\newcommand{\rcam}{{RA/CAM}\xspace}
\newcommand{\imcrypto}{{IMCRYPTO}\xspace}

\definecolor{brightcerulean}{HTML}{1e91d6}
\definecolor{yellowgreen}{HTML}{8fc93a}
\definecolor{meatbrown}{HTML}{e4cc37}
\definecolor{dazzledblue}{HTML}{30638e}
\definecolor{spanishred}{HTML}{bf1a2f}
\definecolor{brightblue}{HTML}{00ccff}

\graphicspath{{images/}}
\newcommand{\bluHL}{\textcolor{black}}

\begin{document}
%\fontsize{9}{12}\selectfont
\title{IMCRYPTO: An In-Memory Computing Fabric \\for AES Encryption and Decryption}

%Copyright information
%Empty

\author{Dayane~Reis,~\IEEEmembership{Student Member,~IEEE,}
        Haoran~Geng,
        Michael~Niemier,~\IEEEmembership{Senior Member,~IEEE}
        and~Xiaobo~Sharon~Hu,~\IEEEmembership{Fellow,~IEEE}% <-this % stops a space
%\thanks{Manuscript received xxxxxx, xxxx; revised xxxxxx, xxxx.}
\thanks{This work was supported in part by ASCENT, one of six centers in JUMP, a Semiconductor Research Corporation (SRC) program sponsored by DARPA. Date of publication xx xx, xxxx; date of current version xx xx, xxxx. (\textit{Corresponding author: Xiaobo Sharon Hu.})}
\thanks{D. Reis, H. Geng, M. Niemier, and X. S. Hu are with the Department
of Computer Science and Engineering, University of Notre Dame, Notre Dame,
IN, 46556, USA. e-mail: shu@nd.edu.}% <-this % stops a space

}

%\author{}

\maketitle

\begin{abstract}
This paper proposes IMCRYPTO, an in-memory computing (IMC) fabric for accelerating AES encryption and decryption. IMCRYPTO employs a unified structure to implement encryption and decryption in a single hardware architecture, with combined (Inv)SubBytes and (Inv)MixColumns steps. Because of this step-combination, as well as the high parallelism achieved by multiple units of random-access memory (RAM) and  random-access/content-addressable memory (RA/CAM) arrays, IMCRYPTO achieves high throughput encryption and decryption without sacrificing area and power consumption. Additionally, due to the integration of a RISC-V core, IMCRYPTO offers programmability and flexibility. IMCRYPTO improves the throughput per area by a minimum (maximum) of 3.3$\times$ (223.1$\times$) when compared to previous ASICs/IMC architectures for AES-128 encryption. Projections show added benefit from emerging technologies of up to 5.3$\times$ to the area-delay-power product of IMCRYPTO.
\end{abstract}


%

%%%%%%%%%%%%%%%%%%%%%%%%%%%%%%%%%%%%%%%%%%%%%%%%%%%%%%%%%%%%%%%%%%%
\section{Introduction}
\label{sec:introduction}
%%%%%%%%%%%%%%%%%%%%%%%%%%%%%%%%%%%%%%%%%%%%%%%%%%%%%%%%%%%%%%%%%%%
 
 Advanced Encryption Standard (AES) \cite{daemen1999aes} is a widely used encryption method used in a variety of computer systems. Despite the excellent security of AES, its software-based implementations can be computationally expensive, resulting in low throughput. High throughput AES encryption/decryption is extremely desirable in many applications (e.g., virtual private networks, electronic financial transactions, etc.). As demonstrated in previous work (e.g., \cite{stevens05, chaves06, ueno20, sayilar14, zhang18,xie18}), hardware accelerators designed for AES encryption/decryption can achieve high throughput while enabling area-efficient design alternatives for resource-constrained environments such as edge computing. AES hardware accelerators are usually application-specific integrated circuits (ASICs) or field-programmable gate array (FPGA)-based co-processors that implement the steps for the AES algorithm, i.e., AddRoundKey, SubBytes/InvSubBytes, ShiftRows, and MixColumns/InvMixColumns. 
 
 The core functions in AES involve mainly simple operations, which makes AES encryption/decryption memory bound. To address the memory bottleneck of AES, in-memory computing (IMC) architectures (i.e., with modified memory cells or peripherals) may be designed to perform operations commonly found in AES-based encryption/decryption (e.g., bit shifts, byte permutations and XOR). IMC-based architectures operate on memory words without the need for data transfers to a processing unit (e.g., \cite{zhang18, xie18}). Despite the potential for reduced data traffic, work in \cite{zhang18, xie18} must make tradeoffs between area/power efficiency and parallelism. To this end, accelerators based on lookup table (LUT) and pre-computed operations (e.g., \cite{stevens05, chaves06}) have demonstrated high levels of parallelism and throughput. In this regard, memories based on emerging technologies (e.g., phase-change memories (PCMs), resistive random-access memories (RRAMs), spin-transfer torque magnetoresistive random-access memories (STT-MRAMs), ferroelectric field-effect transistor-based random-access memories (FeFET-RAMs) can considerably improve the density and static power consumption of IMC architectures (albeit at the expense of longer write latencies) \cite{yu2016emerging, sun2018memory, angizi2019accelerating}.
 
 \bluHL{Besides architecture and technology considerations, security and flexibility in supporting different modes of AES are desirable features of hardware accelerators. For the former, the ability to run new encryption/decryption algorithms that resist emerging attacks is highly desirable \cite{bossuet13}. For the latter, different modes of AES may be selected for different applications, e.g., cipher block chaining (CBC) and the counter (CTR) mode can be used for data streaming and for protecting data storage devices \cite{sasongko21, ege2011memory, weidong05}. As another example, authenticated encryption guarantees both confidentiality (data secrecy) and integrity (data authenticity), which has utility in distinct application spaces and can be performed with Galois/Counter Mode (GCM). Considering this wide range of usage scenarios, having an AES accelerator that supports different modes of AES may reduce costs with system/intellectual property (IP) re-design. As an additional benefit, a general-purpose, programmable accelerator for AES could accommodate other symmetric block ciphers and hashing functions such as the Secure Hash Algorithm (SHA)-256 or the Message Digest Algorithm (MD)-5.}
 
 %Although LUT or memory-based accelerators have great potential for accelerating AES encryption/decryption, the benefits may come at the expense of larger area and higher power consumption. 
 
In this paper, we propose a \bluHL{programmable} in-memory computing fabric for AES acceleration (\imcrypto) that achieves the same parallelism as LUT methods while avoiding large area/power overheads. \imcrypto is based on both conventional and compute-enabled random-access memory (RAM) arrays \cite{aga17}, as well as dual-function random access memory/content adressable memory (\rcam) arrays. Specifically, the \rcam arrays employed in \imcrypto enable encryption and decryption to be performed with a common structure. To improve the computational efficiency, the steps SubByte/MixColumns (for encryption), and InvSubBytes/AddRoundKey/InvMixColumns (for decryption) are combined and executed in one step with our design. The proposed fabric supports pipelining and enables high-throughput execution. The contributions of our work are summarized below:

 \begin{itemize}
     \item We propose \imcrypto, a highly-parallel IMC fabric based on RAM and \rcam arrays that provide high-throughput AES encryption/decryption with a common hardware structure.
    
    \item We propose a compact implementation of the AES algorithm with \imcrypto, which combines certain steps in encryption (and decryption) to reduce computation load.
    
    \item \bluHL{We propose a RISC-V based controller that implements customized instructions in \imcrypto. The flexibility of \imcrypto enable multiple modes of AES and other block ciphers to be implemented.}
    
 \end{itemize}
 
We have performed detailed circuit-level, simulation-based evaluations of our proposed \imcrypto architecture. We consider figures of merit such as delay, area, power, throughput, area-delay-power product (ADPP), and throughput per area and compare them with state-of-the-art ASIC and IMC-based accelerators for AES encryption \cite{ueno20, sayilar14, zhang18, xie18}. Results indicate minimum (maximum) throughput per area improvements of 3.3$\times$ (223.1$\times$) and minimum (maximum) ADPP improvements of 1.2$\times$ (6,238.3$\times$) for CMOS-based \imcrypto when compared to ASIC and IMC accelerators for AES-128 encryption of a 1MB plaintext. (The wide spectrum of improvement is a result of different circuit and design paradigms employed in each accelerator). IMCRYPTO can also leverage emerging memory technologies to further boost the ADPP improvements over a conventional CMOS-based implementation. Our analysis suggests that employing emerging technologies in the design of memory units in \imcrypto could improve the ADPP by an additional factor of up to 5.3$\times$.

\section{Background and Related Work}
\label{sec:background}
%%%%%%%%%%%%%%%%%%%%%%%%%%%%%%%%%%%%%%%%%%%%%%%%%%%%%%%%%%%%%%%%%%%
In this section, we briefly review the basics of AES encryption/decryption, and discuss related work on hardware accelerators for AES encryption/decryption.

\subsection{AES Basics}

Fig. \ref{fig:encryption_decryption}(a) and Fig. \ref{fig:encryption_decryption}(b) illustrate the steps performed in $N$ rounds of AES encryption and decryption. As a variant of the Rijndael block cipher, AES employs a fixed block size of 128 bits, and a key size of 128, 192, or 256 bits (for AES-128, AES-192, or AES-256, respectively). Round keys are derived from the main key for each one of the $N$ rounds using a key schedule algorithm \cite{daemen1999aes}.

The 128-bit block is arranged as a 4$\times$4 array of bytes, which is the state array. Each step in the AES encryption (decryption) takes a state array as an input and produces an output state array that is passed to the next step (or returned as the final encryption or decryption result at the end of $N$ rounds). We represent elements of the input (output) state array as $a_{i,j}$ ($b_{i,j}$), where $i={0,1,2,3}$ and $j={0,1,2,3}$. The steps of AES encryption (decryption) are summarized below.

\subsubsection{AddRoundKey}

In the AddRoundKey step, every byte of a round key $rk$ is combined to the corresponding byte of the input state through an XOR operation, i.e., $b_{i,j}=a_{i,j} \oplus rk_{i,j}$. 

\subsubsection{SubBytes and InvSubBytes}

During SubBytes and InvSubBytes, each byte $a_{i,j}$ in the input state array is replaced by a substitution byte to form the output state array, i.e., $b_{i,j}=Sbox(a_{i,j})$. Since each byte $a_{i,j}$ in the 0x00--0xFF interval maps to an unique substitution byte, encryption/decryption \textit{Sbox} tables can be pre-computed and stored in LUTs (or memory units) to improve the speed of AES. While pre-stored table-based implementations of the SubBytes and InvSubBytes steps may be vulnerable to cache timing attacks in a traditional central processing unit (CPU) architecture, they are well-suited for FPGA or ASIC-based accelerator designs  \cite{eran2010}.

\subsubsection{ShiftRows and InvShiftRows}

During ShiftRows and InvShiftRows, the bytes of the input state array are circularly shifted to the left (right) by a fixed offset amount (0, 1, 2 or 3 positions) depending on the index $i$ of the byte. The output state is then formed by the byte-shifted input state. For $i=0$, no circular byte shifts are performed ($b_{i,j}=a_{i,j}$). For $i=1$, $2$, and $3$, the offsets of the circular byte shifts are defined as $1$, $2$, and $3$, respectively, with a left (right) direction for encryption (decryption). 

 \begin{figure}[t]
  \centering
  \includegraphics[scale=0.25]{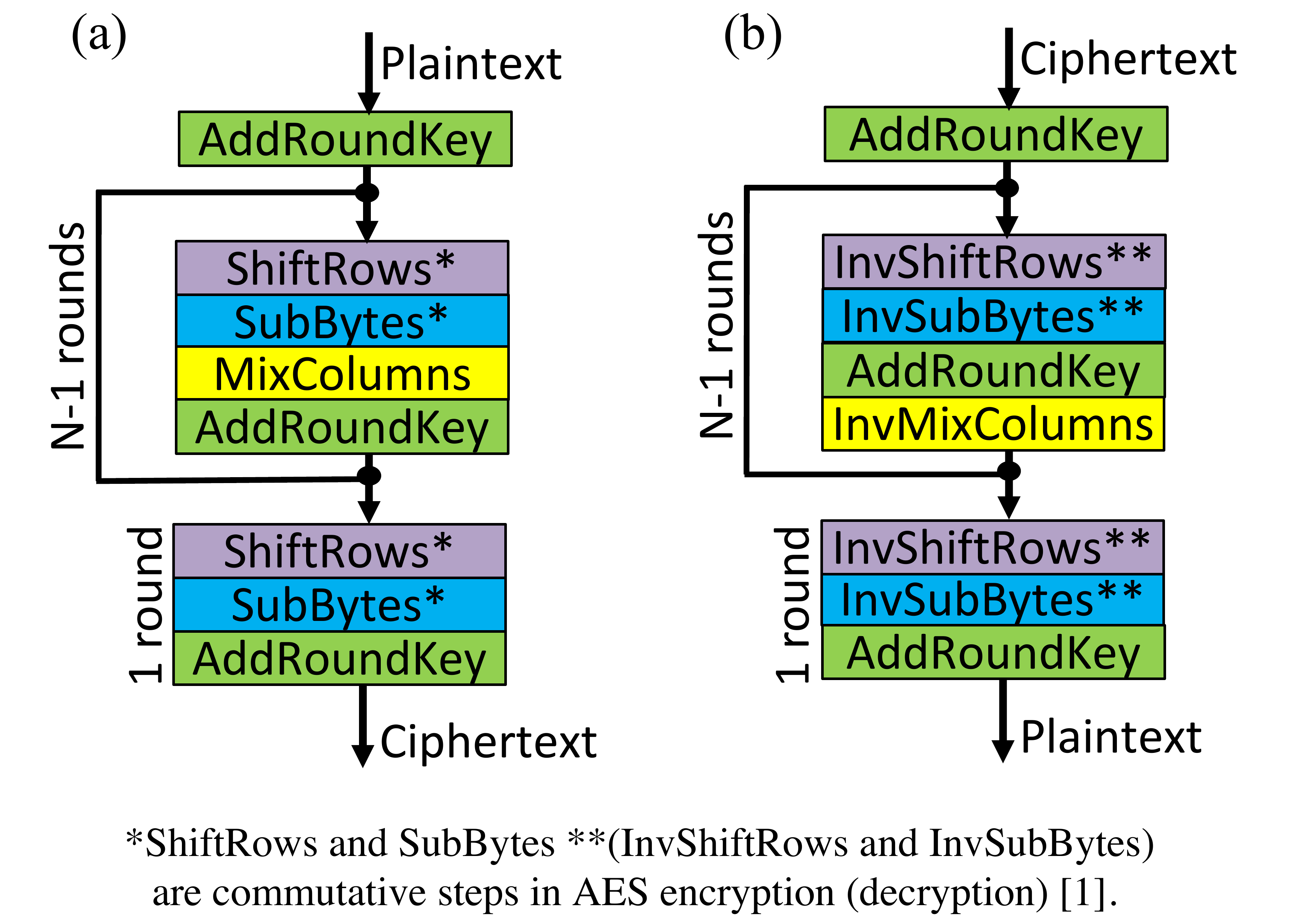}
  \vspace{-1ex}
  \caption{The steps performed in the $N$ rounds of AES (a) encryption and (b) decryption. The values of $N$ are 10 (for AES-128), 12 (for AES-192), and 14 (for AES-256). (Inv)ShiftRows and (Inv)SubBytes are commutative steps \cite{daemen1999aes}. }
  \label{fig:encryption_decryption}
    \vspace{-2ex}
\end{figure}

%  \begin{figure}[t]
%   \centering
%   \includegraphics[scale=0.25]{Figures/addroundkey.pdf}
%   \vspace{-3ex}
%   \caption{The AddRoundKey step.}
%   \label{fig:addroundkey}
%     \vspace{-2ex}
% \end{figure}

 \begin{figure}[t]
  \centering
  \includegraphics[scale=0.25]{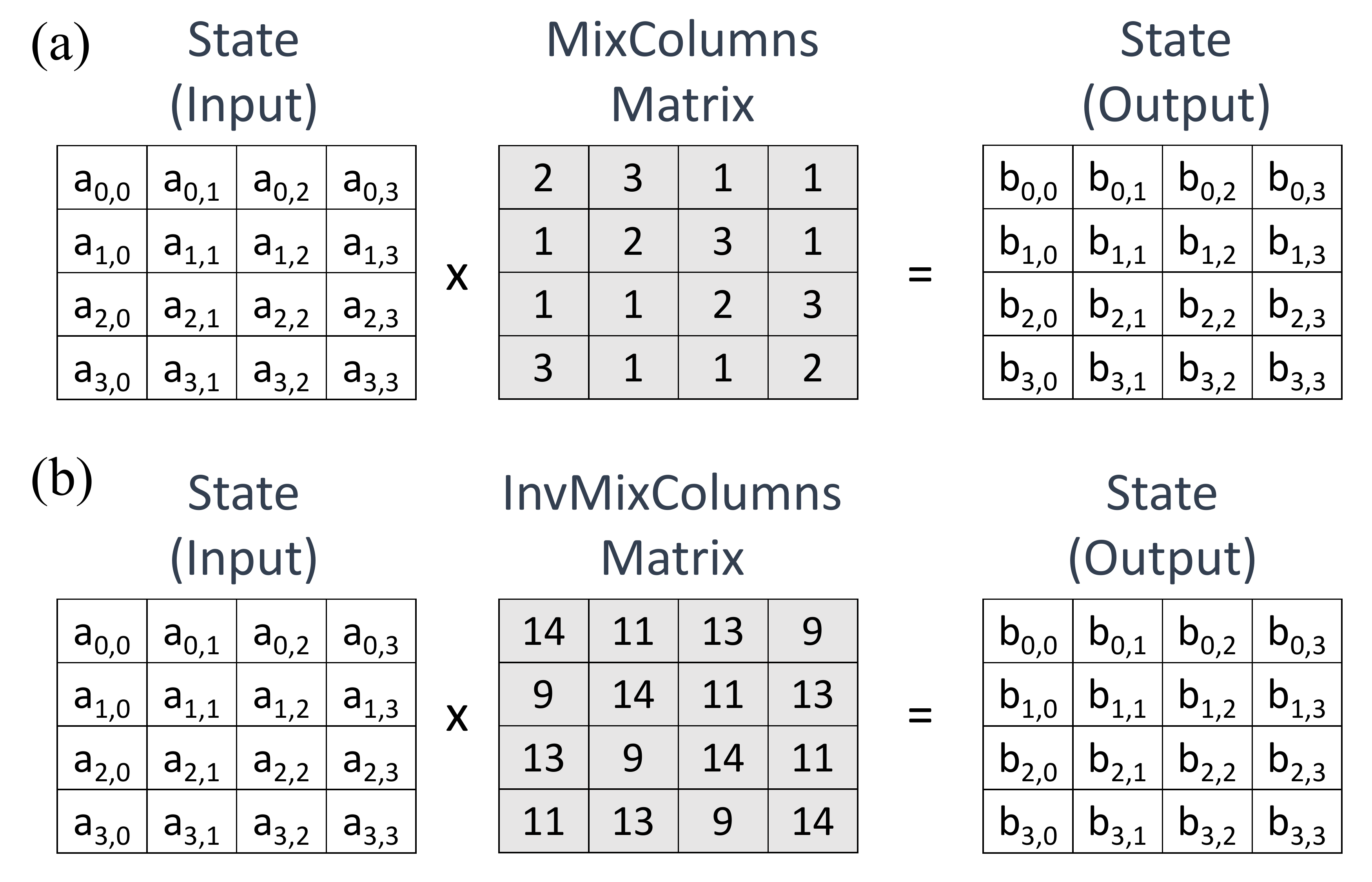}
  \vspace{-2ex}
  \caption{The (a) MixColumns and (b) InvMixColumns steps of AES-based encryption and decryption, respectively.}
  \label{fig:mixcolumns}
    \vspace{-2ex}
\end{figure}

\subsubsection{MixColumns and InvMixColumns}

During MixColumns and InvMixColumns, a linear transformation is applied to the input state matrix to form the output state matrix. This transformation consists of a matrix multiplication over the Galois Field ($GF(2^8)$) between a fixed matrix and the input state matrix (see Fig. \ref{fig:mixcolumns}). %Matrix multiplication over the $GF(2^8)$ boils down to a series bit shifts and XOR operations, which can be implemented with specialized circuits in hardware accelerators. However, s
%Since a fixed matrix is used, i
As with SubBytes (InvSubBytes), it is possible to speed-up the MixColumns (InvMixColumns) steps by pre-computing the products between every byte value in the range 0x00--0xFF and the multiplication coefficients. 

\subsubsection{Modes of Operation}

\bluHL{Modes of operation ensure confidentiality and authenticity when a cipher's single-block operation encrypts/decrypts an amount of data larger than a 128-bit block. Different modes of operation are commonly used in addition to the basic version of the AES algorithm, \textit{a.k.a.} the ECB (Electronic Code Book). For instance, the CBC, the CFB (Cipher FeedBack), the OFB (Output FeedBack), the CTR, and the GCM are AES modes that require simple operations (such as XOR or addition) to be performed between the state array and randomly-generated numbers (or counter values) at the start of every AES round. (For details on different modes of AES and their practical use, see \cite{blazhevski2013modes, almuhammadi2017comparative}).}

\subsection{Related Work}
\label{sec:related_work}

Numerous research efforts have been devoted to developing FPGA or ASIC-based hardware to accelerate AES (e.g., \cite{chaves06, stevens05, sayilar14, zhang18, ueno20, xie18, li05, mathew11}). That said, designing a hardware accelerator for AES that can provide high throughput without large overheads in terms of area and power consumption can be a challenging task. Area-efficient AES accelerators that provide high throughput are highly sought-after intellectual property (IP) blocks for resource-constrained environments (e.g., in embedded systems \cite{ritambhara17}).

One of the strategies to minimize area without compromising performance is to re-use circuit blocks and to design shared datapaths for encryption and decryption \cite{wolker02, ueno20}. In this regard, \cite{ueno20} introduces an architecture for AES with shared encryption/decryption datapaths. Although their shared-datapath strategy can help reduce area, some circuits used for encryption and decryption still need to be implemented separately in this architecture. Furthermore, the design proposed in \cite{ueno20} is a standalone ASIC, which requires plaintexts and ciphertexts to be fetched/stored in memory units. As we show in our evaluation (see Sec. \ref{sec:evaluation}), the latency and energy overheads from data transfers can significantly impact the performance of hardware accelerators for AES. 
%Alternatively, our proposed \imcrypto is an IMC-based accelerator that executes operations with compute-enabled arrays and customized memory peripherals. Our design enables the SubBytes/InvSubBytes and MixColumns/InvMixColumns steps of AES \textit{encryption and decryption} to be performed with a common hardware structure, and in a combined fashion. 

\bluHL{Reducing the impact of data transfers on a system's performance is a desired feature for AES accelerators. To this end, IMC could potentially alleviate the burden of data transfers between the memory and the accelerator. Reference \cite{zhang18} proposes a reconfigurable cryptographic processor that uses IMC and near-memory computing for supporting AES and elliptic curve cryptography (ECC). Even though their IMC-based design improves the performance, it lacks the flexibility of supporting multiple modes of AES and other cryptographic algorithms (e.g., SHA-256). In this regard, reference \cite{sayilar14} proposes a highly reconfigurable ASIC cryptographic processor. It can support most of the existing cryptographic algorithms and potentially can support future algorithms. Although their throughput per area is much higher than the best CPU and graphic processing unit (GPU), they do not account for data movement in their evaluation (similar to \cite{ueno20}).}

\bluHL{Reference \cite{xie18} presents another IMC accelerator for AES based on PCMs, which is intended to protect the contents of the non-volatile, off-chip main memory in case an attacker gets physical access to it. The accelerator can simultaneously encrypt plaintext within each memory bank, and the entire encryption process can be completed without exposing the results to the memory bus. However, in a usage scenario where this IMC accelerator is used to perform decryption and send the resulting plaintext to the CPU, the architecture may be vulnerable to side channel attacks as plaintext sent from the off-chip memory would be exposed to the memory bus.  The design also requires expensive writes to the non-volatile PCM (including results from the SubBytes and MixColumns steps of AES), which is a concern due to the high write latency and high write energy of the PCM memory \cite{yu2016emerging}.}

\section{\imcrypto Fabric}
\label{sec:cim_design}
%%%%%%%%%%%%%%%%%%%%%%%%%%%%%%%%%%%%%%%%%%%%%%%%%%%%%%%%%%%%%%%%%%%

\begin{figure}[t]
  \centering
  \includegraphics[scale=0.23]{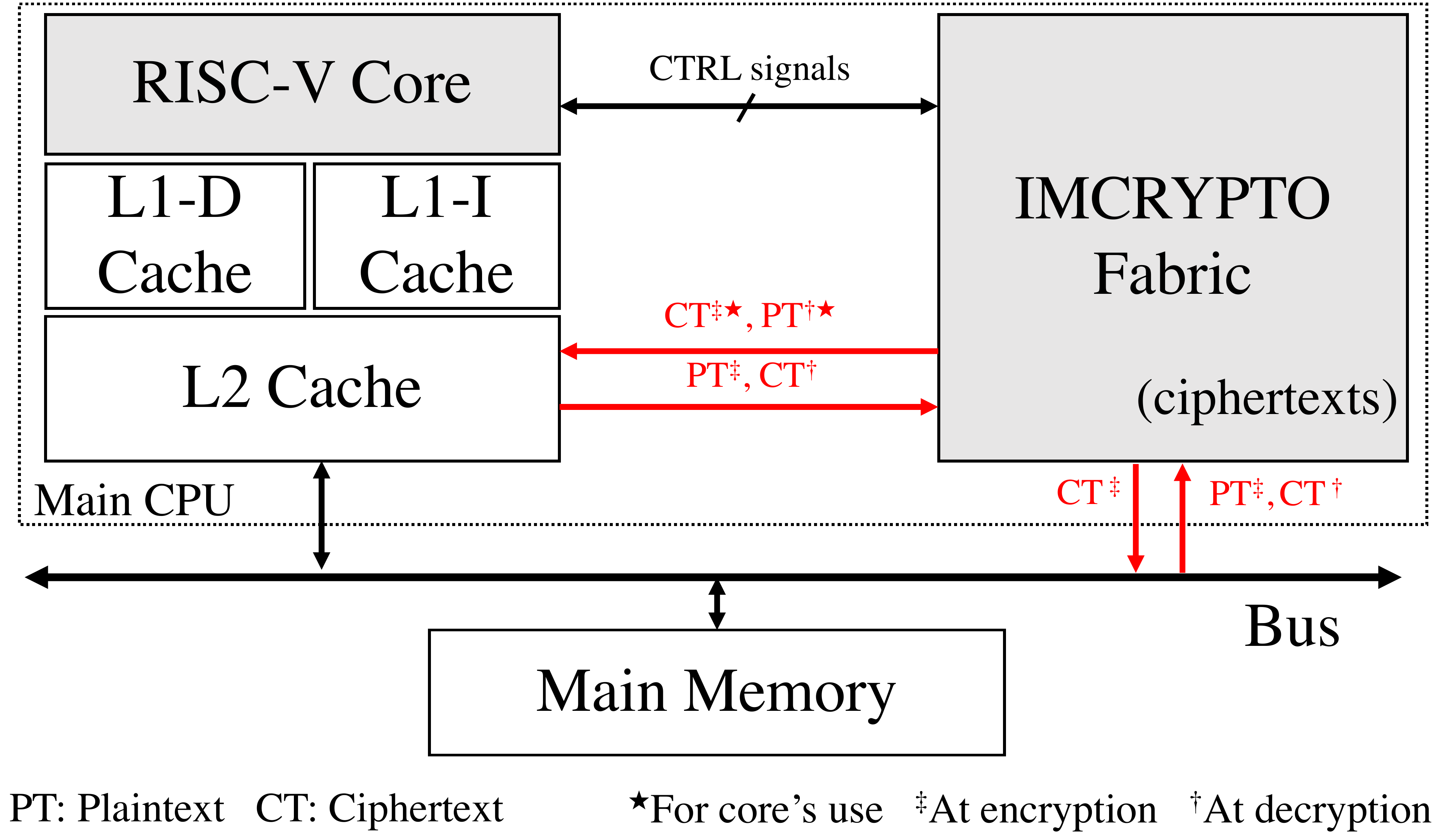}
  \caption{An overview of an \imcrypto-based system.}
  \label{fig:high_level}
    \vspace{-2ex}
   
  % \begin{flushleft}
%{\bf Notes:} $\mathbf{{}^\star}$Bi-directional data flow is permitted between the CPU L2 cache and the \imcrypto fabric.\\
%\end{flushleft}
\end{figure}

%%%%%%%%%%%%%%%%% FROM SEC. 3
 \begin{figure*}[t]
  \centering
  \includegraphics[scale=0.20]{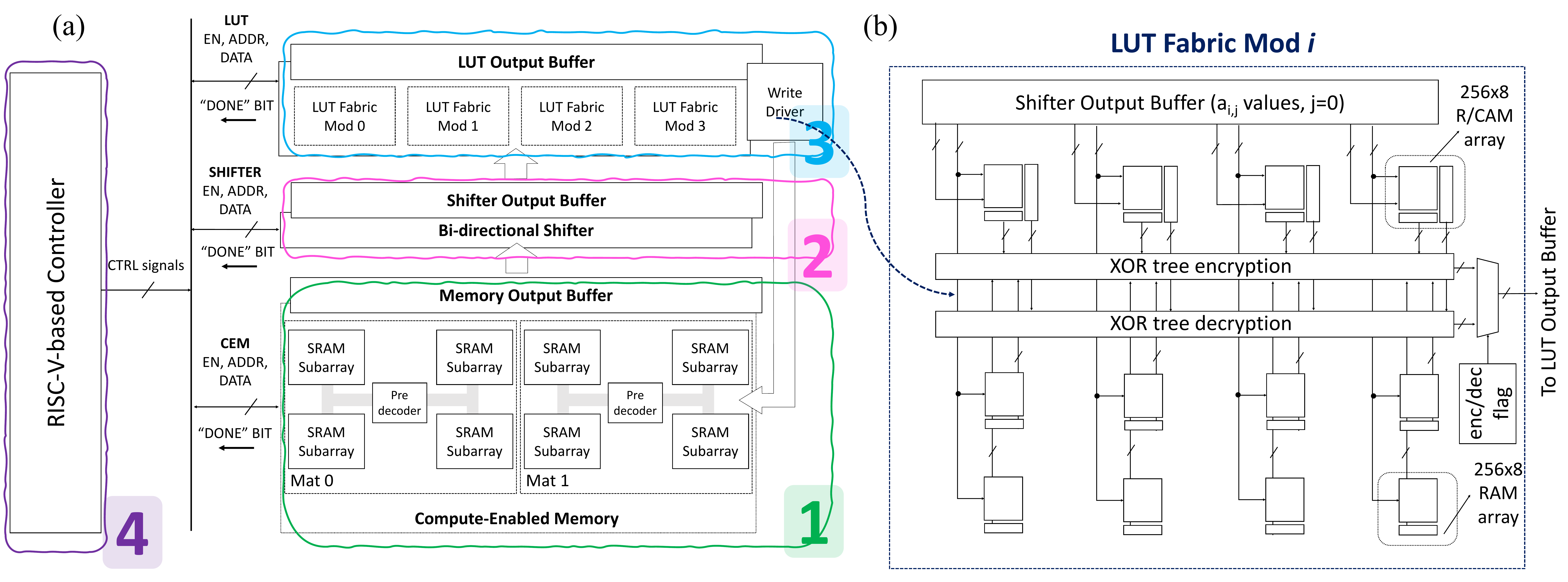}
  \vspace{-2ex}
  \caption{(a) A high-level view of the \imcrypto accelerator, (b) with details of 1 LUT fabric module. }
  \label{fig:architecture}
    \vspace{-2ex}
\end{figure*}
%%%%%%%%%%%%%%%%%

\bluHL{We propose the \imcrypto fabric for AES encryption/decryption. Our proposed in-memory accelerator is placed on the same chip as the CPU, at the level of the last level cache (LLC). Placing \imcrypto on-chip and at the level of the LLC facilitates data exchange with the CPU and creates a mechanism for storing ciphertexts inside \imcrypto's internal memory \textbf{(i.e., \imcrypto works as the LLC for secure data)}. Furthermore, data is properly protected before being sent to the main memory through the memory bus. \imcrypto employs a light-weight, general-purpose core to program \imcrypto's functionality. For simplicity, we use a RISC-V core in \imcrypto, but other cores can also be designed for this purpose. Thanks to the programmability enabled by the RISC-V core, \imcrypto offers the needed flexibility at the application level (as mentioned in Sec. \ref{sec:introduction}).}

\bluHL{Note that using \imcrypto as the LLC for secure data should not significantly impact the performance of the main CPU for executing non-secure workloads, since the L1/L2 caches can have their content accessed directly by the RISC-V core without going through \imcrypto. Furthermore, with \imcrypto working simultaneously as an accelerator for AES encryption/decryption and the LLC for secure data, the L1 and L2 caches (represented in Fig. \ref{fig:high_level}) are reserved to the storage of non-sensitive data. Thus, they can freely transfer data from/to the main memory through the bus and from/to the RISC-V core without any negative impact to the security.}

\bluHL{Fig. \ref{fig:high_level} provides an overview of the \imcrypto-based system, where the interaction between \imcrypto and the other components of a computer system (i.e., the main memory and L1/L2 caches) is explicitly shown. %Note that, in our architecture, \imcrypto works as both an IMC accelerator for AES encryption/decryption and a LLC for secure data. 
Encryption/decryption requests to \imcrypto can be made by the RISC-V core, which communicates with \imcrypto through a set of control (CTRL) signals. The RISC-V uses the customized instruction ``TEXT'' (to be described in Sec. \ref{sec:riscv}, Table \ref{tab:instructions_type_i}) to pass the address of a plaintext (PT) or a ciphertext (CT) for encryption/decryption with \imcrypto. Round keys are generated on-the-fly by the RISC-V core and directly provided to \imcrypto. Below, we describe how \imcrypto processes encryption/decryption requests by the RISC-V core:}

\begin{itemize}
    \item \bluHL{\textbf{At encryption}, the instruction ``TEXT'' is used to pass the data address of a plaintext (PT) to \imcrypto. The PT is fetched to (and \textit{temporarily} stored in) the internal memory of \imcrypto from either the L2 cache (in case of a cache hit) or the main memory (in case of a cache miss). After the data transfer process, the PT is encrypted by the IMC infrastructure in \imcrypto, generating a ciphertext (CT). Once encryption finishes, the resulting CT replaces the initial PT in the internal memory of \imcrypto, where it is stored for future use by the RISC-V core (i.e., \imcrypto works as the LLC for secure data). Once the capacity of the \imcrypto's memory is reached (or if there are conflict misses), selected CTs are transferred to main memory via the bus.}
    
    \item \bluHL{\textbf{At decryption}, the instruction ``TEXT'' is used to pass the data address of a CT to \imcrypto. The CT is fetched to the internal memory of \imcrypto from either the L2 cache (in case of a cache hit) or the main memory (in case of a cache miss). After the data transfer process, the CT is decrypted by the IMC infrastructure in \imcrypto, generating a plaintext (PT). Once decryption finishes, the post-decryption PT is not kept in the internal memory of \imcrypto (which is reserved for secure data), but rather is transferred to the CPU's L2 cache to be used by processes running on the CPU.}
\end{itemize}

%The communication between the RISC-V core and \imcrypto is realized by a set of control (CTRL) signals. When performing encryption, the plaintext to be encrypted is fetched from the CPU L2 cache and \textit{temporarily} stored in the internal memory of \imcrypto, alongside the encryption round key. Once encryption finishes, the resulting ciphertext replaces the initial plaintext in  the internal memory of \imcrypto, where it is securely stored. The ciphertexts stored in the \imcrypto fabric can be accessed by the RISC-V core when running applications (e.g., data streaming tasks) through the cache hierarchy. Once the capacity of the \imcrypto's memory is reached (or if there are conflict misses), selected ciphertexts will be transferred to main memory via the bus. Decryption takes place internally in the \imcrypto fabric, and plaintext is not exposed to the memory bus. The post-decryption plaintext generated by \imcrypto is not kept in the internal memory of \imcrypto, but rather is transferred to the CPU L2 cache to be used by processes running on the CPU.

The \imcrypto fabric (\bluHL{Fig. \ref{fig:architecture}(a)}) consists of circuits that may be grouped into four distinct blocks. The blocks, numbered from 1 to 4 in Fig. \ref{fig:architecture}(a), \bluHL{are (1) the compute-enabled memory (CEM), (2) the bi-directional shifter, (3) the LUT fabric (comprised of 4 modules), and (4) the RISC-V based controller}. The circuits in each block as well as the communication between them are described below.

\subsection{The Compute-Enabled Memory}
\label{sec:ce_memory}

The AddRoundKey step of AES encryption (decryption) is the addition of a round key to a plaintext (ciphertext) through a simple XOR operation. \bluHL{\textbf{To enable the AddRoundKey step in AES encryption (and in the initial round of AES decryption) the proposed architecture implements the compute-enabled memory (CEM) block (block 1 in Fig. \ref{fig:architecture}(a)).}} \bluHL{The CEM of \imcrypto employ circuits similar to those in \cite{reis19, aga17} --- i.e., customized sense amplifiers --- which enable bitwise Boolean logic between two memory words stored in a memory array.}

\bluHL{The CEM receives plaintext from the CPU L2 cache, performs logic operations on the AES state array, and stores the resulting ciphertext after encryption. The encrypted data in \imcrypto can be directly read from the CEM by the CPU cache structure and sent to external parties using communication protocols like the peer-to-peer (P2P) or Attribute-Based Encryption (ABE) (used in IoT systems). Another usage scenario for direct access to the CEM is when the CPU core performs operations on data that has been previously encrypted and stored in \imcrypto. In this case, \imcrypto allows for the data to be decrypted prior to sending it to be used by the processes running on the CPU core.} %By placing \imcrypto on the same chip as the CPU instead of the main memory (as with \cite{xie18}), we avoid exposing the plaintext at the memory bus, which could make the system susceptible to side-channel attacks.

As an on-chip memory, the size of a CEM needs to be within the acceptable limits for a CPU cache (i.e., from tens of KB up to a few MB). \bluHL{A larger CEM enables more ciphertexts to be stored in \imcrypto, which can be useful when cached ciphertexts are sent to external parties using communication protocols.} On the other hand, bigger memories have longer access times and consume more power. To allow for a compromise between memory size, access times and energy consumption, the CEM block of \imcrypto is a 1 MB memory that consists of a tiled, compute-enabled SRAM structure \cite{aga17}. With the aid of customized sense amplifiers, the CEM can execute XOR operations between two memory words without the need for reading the data out to an external processing unit.% Importantly, the AddRoundKey step needed in the remaining rounds of AES decryption is performed by customized encoders rather than (see Sec. \ref{sec:ram_cam_module} and Sec. \ref{sec:merged_steps} for more details).

\subsection{The Bi-Directional Shifter Block}
\textbf{The bi-directional shifter block (labeled as \bluHL{block 2} in Fig. \ref{fig:architecture}(a)) performs the byte permutations needed by the ShiftRows (InvShiftRows) steps of AES encryption (decryption).} Internally, the circuit of the bi-directional shifter block is similar to that described in \cite{mathew11}.

\subsection{The LUT Fabric}
\label{sec:ram_cam_module}

\textbf{The LUT fabric (labeled as \bluHL{block 3} in Fig. \ref{fig:architecture}(a)) is a critical block in \imcrypto  as it executes the SubBytes/InvSubBytes and MixColumns/InvMixColumns (the most time-consuming steps of AES).} With the use of small memory elements (i.e., 6T-SRAM and \rcam arrays of size 256$\times$8) and customized peripherals, it is possible to perform these steps in a highly parallel and combined fashion with low latency. The size of the memory elements (i.e., 256$\times$8) is chosen so it is possible to store all the possible 256 pre-computed bytes of combined SubBytes and MixColumns.

The LUT fabric block comprises 4 LUT fabric modules that execute 1 matrix multiplication in GF($2^8$). These four identical LUT fabric modules in the LUT fabric block operate \textit{in parallel} to compute each column of the output state array (i.e., $j=0$,$1$,$2$,$3$) in the MixColumns/InvMixColumns and SubBytes/InvSubBytes steps. The internal structure of a LUT fabric module is depicted in Fig. \ref{fig:architecture}(b). Each LUT fabric module consists of (i) 4 \rcam arrays with customized peripherals, (ii) 8 RAM arrays, and (iii) 2 sets of cascaded XOR gates (\textit{a.k.a.} XOR trees, which are employed to perform additions over $GF(2^8)$). The number of memory elements of each type inside a LUT fabric module allows combined steps to be performed without serializing operations. More details about the execution of combined steps will be given in Sec. \ref{sec:merged_steps}.

%%%%%%%%%%%%%%%%%%%%%%%%%%%%%%%%%%%%%%%%%%%%%%%%%%%%%%%%%%

\begin{table} [t]
\setlength\extrarowheight{2pt}
\caption{Customized RISC-V instructions designed for \imcrypto}
\begin{subtable}{1\columnwidth}
\centering

\scalebox{.90}{\begin{tabular}{|c|c|c|c|c|l|}
\hline
\multicolumn{2}{|c|}{\multirow{2}{*}{Instruction}} & \multicolumn{2}{c|}{Function Code} & \multirow{2}{*}{Assembly}                                  & \multicolumn{1}{c|}{\multirow{2}{*}{Description}}                                                  \\ \cline{3-4}
\multicolumn{2}{|c|}{}                             & Bits 1-7                  & Bit 8  &                                                            & \multicolumn{1}{c|}{}                                                                              \\ \hline
\multirow{2}{*}{TEXT}    & L                       & \multirow{2}{*}{0000000}  & 0      & \begin{tabular}[c]{@{}c@{}}TEXT\\ address, $s1$\end{tabular}    & \begin{tabular}[c]{@{}l@{}}Load 128 bits from \\ \imcrypto to RISC-V\\registers during decryption\end{tabular}        \\  \cline{2-2} \cline{4-6} 
                         & S                       &                           & 1      & \begin{tabular}[c]{@{}c@{}}TEXT\\ $s1$, address\end{tabular}    & \begin{tabular}[c]{@{}l@{}}Store 128 bits to \\ \imcrypto from RISC-V\\registers during encryption\end{tabular}       \\ \hline

\multirow{2}{*}{SFTR}    & E                       & \multirow{2}{*}{0000100}  & 0      & \begin{tabular}[c]{@{}c@{}}SFTR\\ $s1$ \end{tabular}    & \begin{tabular}[c]{@{}l@{}}Trigger ShiftRows\\ on $s1$ (encryption)\end{tabular}                           \\ \cline{2-2} \cline{4-6} 
                         & D                       &                           & 1      & \begin{tabular}[c]{@{}c@{}}SFTR\\ $s1$ \end{tabular}    & \begin{tabular}[c]{@{}l@{}}Trigger InvShiftRows\\ on $s1$ (decryption)\end{tabular}                        \\ \hline
\multirow{2}{*}{SUBMX}   & E                       & \multirow{2}{*}{0001000}  & 0      & \begin{tabular}[c]{@{}c@{}}SUBMX\\ $s1$\end{tabular}    & \begin{tabular}[c]{@{}l@{}}Trigger combined \\SubBytes+\\ MixColumns \\(encryption) on $s1$\end{tabular}       \\ \cline{2-2} \cline{4-6} 
                         & D                       &                           & 1      & \begin{tabular}[c]{@{}c@{}}SUBMX\\ $s1$\end{tabular}    & \begin{tabular}[c]{@{}l@{}}Trigger combined \\InvSubBytes+ \\InvMixColumns\\ (decryption) on $s1$\end{tabular} \\ \hline
\multirow{2}{*}{SBOX}    & E                       & \multirow{2}{*}{0010000}  & 0      & \begin{tabular}[c]{@{}c@{}}SBOX\\ $s1$\end{tabular}    & \begin{tabular}[c]{@{}l@{}}Trigger SubBytes\\ (last round\\ of encryption) on $s1$\end{tabular}              \\ \cline{2-2} \cline{4-6} 
                         & \multicolumn{1}{l|}{D}  &                           & 1      & \begin{tabular}[c]{@{}c@{}}SBOX\\ $s1$\end{tabular}    & \begin{tabular}[c]{@{}l@{}}Trigger InvSubBytes\\ (last round\\ of decryption) on $s1$\end{tabular}           \\ \hline
\end{tabular}}
\vspace{1ex}
\caption{RV32I I-type Instructions for \imcrypto}
\label{tab:instructions_type_i}
\end{subtable}
\par
%\vspace{2ex}
%%%%%%%%%

\begin{subtable}{1\columnwidth}
\centering
%\scriptsize
\scalebox{0.95}{\begin{tabular}{|c|c|c|c|c|}
\hline
\multirow{2}{*}{Instruction} & \multicolumn{2}{c|}{Function Code} & \multirow{2}{*}{Assembly}                                  & \multirow{2}{*}{Description}                                                                                  \\ \cline{2-3}
                             & Bits 1-7        & Bits 8-10        &                                                            &                                                                                                               \\ \hline
IMMOVE                       & 0000000         & 000              & \begin{tabular}[c]{@{}c@{}}IMMOVE\\ $s1$, $sd$\end{tabular}    & \begin{tabular}[c]{@{}c@{}}Move data in memory\\ from address $s1$ to $sd$\end{tabular}                           \\ \hline
IMADD                        & 0000000         & 001              & \begin{tabular}[c]{@{}c@{}}IMADD\\ $s1$, $s2$, $sd$\end{tabular} & \begin{tabular}[c]{@{}c@{}}Store the value of\\ $s1+ s2$ into $sd$\end{tabular}                                    \\ \hline
IMAND                        & 0000000         & 010              & \begin{tabular}[c]{@{}c@{}}IMAND\\ $s1$, $s2$, $sd$\end{tabular} & \begin{tabular}[c]{@{}c@{}}Store the value of\\ $s1$ AND $s2$ into $sd$\end{tabular}                                \\ \hline
IMOR                         & 0000000         & 011              & \begin{tabular}[c]{@{}c@{}}IMOR\\ $s1$, $s2$, $sd$\end{tabular}  & \begin{tabular}[c]{@{}c@{}}Store the value of\\ $s1$ OR $s2$ into $sd$\end{tabular}                                 \\ \hline
IMXOR                        & 0000000         & 100              & \begin{tabular}[c]{@{}c@{}}IMXOR\\ $s1$, $s2$, $sd$\end{tabular} & \begin{tabular}[c]{@{}c@{}}Store the value of\\ $s1$ XOR $s2$ into $sd$\end{tabular}                                \\ \hline
IMNOT                        & 0000000         & 101              & \begin{tabular}[c]{@{}c@{}}IMNOT\\ $s1$, $sd$\end{tabular}     & \begin{tabular}[c]{@{}c@{}}Store the value of\\ $\overline{s1}$ into $sd$\end{tabular}                       \\ \hline
IMCSR                        & 0000000         & 110              & \begin{tabular}[c]{@{}c@{}}IMCSR\\ $s1$, $s2$, $sd$\end{tabular} & \begin{tabular}[c]{@{}c@{}}Circular shift of $s2$ by \\ $s1$ bits to the right,\\ store result in $sd$\end{tabular} \\ \hline
IMSR                         & 0000000         & 111              & \begin{tabular}[c]{@{}c@{}}IMSR\\ $s1$, $s2$, $sd$\end{tabular}  & \begin{tabular}[c]{@{}c@{}}Shift of $s2$ by \\ $s1$ bits to the right,\\ store result in $sd$\end{tabular}          \\ \hline
IMCSL                        & 1000000         & 000              & \begin{tabular}[c]{@{}c@{}}IMCSL\\ $s1$, $s2$, $sd$\end{tabular} & \begin{tabular}[c]{@{}c@{}}Circular shift of $s2$ by \\ $s1$ bits to the left,\\ store result in $sd$\end{tabular}  \\ \hline
IMSL                        & 1000000         & 001              & \begin{tabular}[c]{@{}c@{}}IMSL\\ $s1$, $s2$, $sd$\end{tabular} & \begin{tabular}[c]{@{}c@{}}Shift of $s2$ by \\ $s1$ bits to the left,\\ store result in $sd$\end{tabular}  \\ \hline

\hline
\end{tabular}}
\vspace{1ex}
\caption{RV32I R-type Instructions for \imcrypto}
\label{tab:instructions_type_r}
\end{subtable}
\vspace{-4ex}
\label{tab:instructions}
\end{table}

%%%%%%%%%%%%%%%%%%%%%%%%%%%%%%%%%%%%%%%%%%%%%%%%%%%%%%%%%%

\subsection{RISC-V based Controller}
\label{sec:riscv}

\bluHL{Encryption and decryption in AES requires that certain computational steps are performed in a certain order over $N$ rounds (see Fig. \ref{fig:encryption_decryption}). Moreover, different modes of AES have different data flows and may require additional operations. \textbf{\imcrypto employs a RISC-V based controller (\bluHL{block 4} in Fig. \ref{fig:architecture}(a)) to perform the different steps of AES and support different modes of operation.} Additionally, the RISC-V based controller makes it possible to extend the use of the \imcrypto fabric for other cryptographic algorithms that require similar operations as AES.}

% \begin{figure}[t]
%   \centering
%   \includegraphics[scale=0.35]{Figures/instruction_type.pdf}
%   \vspace{-2ex}
%   \caption{The (a) IMCRYPTO customized instructions RV32I I-type
%  and (b) IMCRYPTO customized instructions RV32I R-type.}
%   \label{fig:type}
%     \vspace{-2ex}
% \end{figure}

\bluHL{To support \imcrypto operations, we introduce 8 RV32I I-type (Table \ref{tab:instructions_type_i}) and 10 RV32I R-type (Table \ref{tab:instructions_type_r}) customized instructions in the RISC-V instruction set architecture (ISA). Two RV32I I-type instructions are used to load/store from/to \imcrypto's CEM (discussed in Sec. \ref{sec:ce_memory}) to/from registers in the RISC-V core. The rest of the I-type instructions are used to trigger each step in AES encryption/decryption. After decoding these customized I-type instructions, the RISC-V based controller generates three main inputs to each block of \imcrypto --- 1 enable (EN) bit, 1 memory address (ADDR), and 1 piece of data (4 registers of 32 bits each). When \imcrypto finishes executing an instruction, it sends back a ``DONE” signal to the RISC-V based controller, which then fetches the next instruction.}

\bluHL{The 10 RV32I R-type customized instructions are used to trigger general-purpose bitwise and arithmetic in-memory operations, such as IMAND, IMOR, IMXOR, IMADD, etc. inside the CEM in \imcrypto. These instructions are used to realize the arithmetic and logic operation inside the memory (IMXOR is also used to achieve the AddRoundKey step of AES). As the in-memory customized R-type instructions are decoded, 1 enable bit (EN), 2 source memory addresses and 1 destination memory address (ADDR) are generated. The CEM operates on the two source memory addresses specified by instructions, performs the corresponding in-memory operation, and stores the result in the destination memory address. The aforementioned control signals (for both RV32I I-type and R-type instructions) are showed in Fig. \ref{fig:architecture}(a).}

\bluHL{We use the opcodes 0000111 and 1000111 to differentiate the regular RISC-V instructions from our customized instructions I-type and R-type. Furthermore, as \imcrypto's instructions are executed inside the CEM, the arithmetic logic unit (ALU) and memories inside the RISC-V processor can be simultaneously used to execute regular RISC-V instructions running in different threads. For example, \imcrypto can use one RISC-V thread to receive data. By employing this multi-threaded technique, the RISC-V core acts as a controller for our \imcrypto accelerator, while it is also available to be used as a general purpose processor.}

\bluHL{By supporting different in-memory operations via the customized RISC-V R-type instructions, \imcrypto can achieve various modes of operation for AES without any hardware modifications. For instance, the CTR mode of AES leverages a counter value which can be stored inside the CEM of \imcrypto. The counter can then be implemented with the use of IMADD operations. As another example, the CBC mode of AES for processing streaming data can be achieved by an additional IMXOR operation between the output values of two consecutive AES 128-bit blocks.}

\bluHL{In addition to the different modes of AES, IMCRYPTO can also support other block ciphers and hash algorithms such as Blowfish, CAST-128, SHA-256 and MD5, which has similar computation operations and complexity as AES. By implementing these algorithms in IMCRYPTO, we can take advantage of high parallelism, high throughput, low data transfer time, and low energy consumption offered by IMC. In Sec. \ref{sec:block_ciphers_example}, we present a use case for an \imcrypto-based implementation of the SHA-256 hashing algorithm. }

\section{\rcam Design}
\label{sec:array}

 \begin{figure}[t]
  \centering
  \includegraphics[scale=0.19]{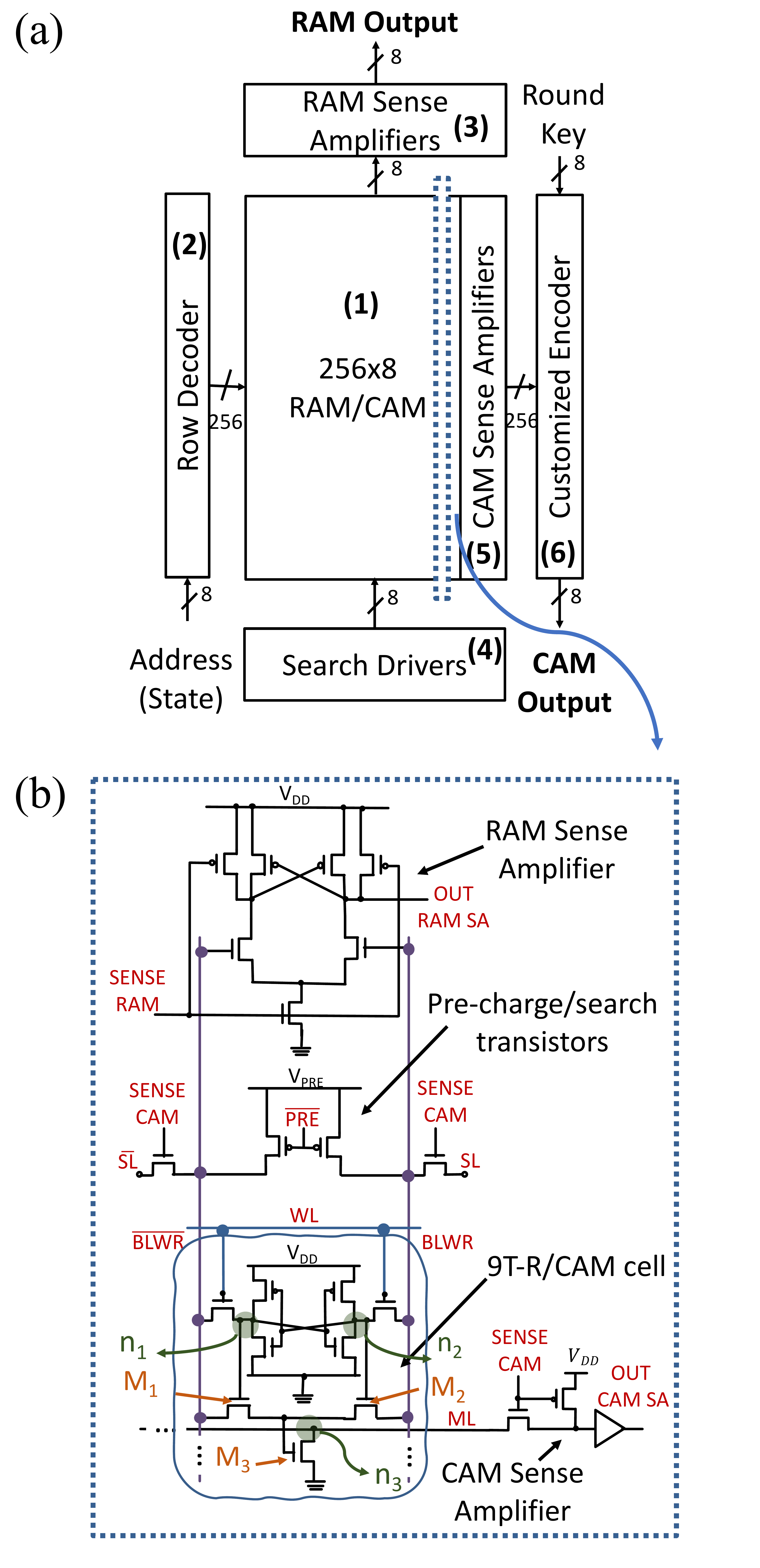}
  \vspace{-1ex}
  \caption{(a) A high level view of the \rcam array. (b) The schematic view of 1 column of the array (with 1 \rcam cell included).}
  \label{fig:array}
    %\vspace{-5ex}
\end{figure}

 \begin{figure*}[!t]
  \centering
  \includegraphics[scale=0.14]{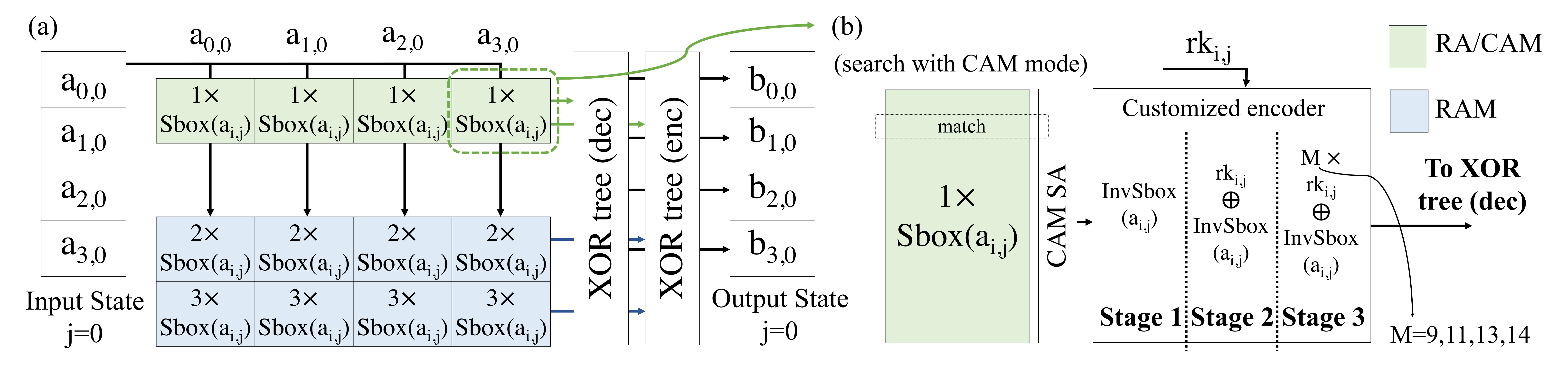}
  \vspace{-2ex}
  \caption{The combined (a) SubBytes and MixColumns steps for AES encryption, and (b) InvSubBytes, AddRoundKey, and InvMixColumns steps for AES decryption.}
  \label{fig:merged}
    %\vspace{-4ex}
\end{figure*}

Here, we introduce the \rcam design in \imcrypto. Fig. \ref{fig:array}(a) provides a high-level view of the components of the \rcam design: (1) the 256$\times$8 \rcam array, (2) the row decoder, (3) the RAM sense amplifiers (SAs), (4) the search drivers, (5) the CAM sense amplifiers (SAs), and (6) the customized encoder. 

Our \rcam design is based on complementary metal–oxide–semiconductor (CMOS) static random-access memory (SRAM) memory cells. While 6T-SRAM cells can be employed for the design of RAM modules (e.g.,\cite{si19}), our \rcam cell design employs 9 transistors (similar to the cell proposed in \cite{li05}). Although \cite{li05} emphasizes the usage of 9T-SRAM cells in the CAM setting only, in \imcrypto we introduce peripheral circuits (i.e., a row decoder and RAM sense amplifiers) that enables the array to work as both a RAM or a CAM (hence, a \rcam array is comprised of 9T-\rcam cells). The CAM functionality of the \rcam array enables both the SubBytes and InvSubBytes steps to be performed with a common structure for encryption and decryption.

The schematic view of one column of the \rcam array is shown in Fig. \ref{fig:array}(b). Although we only show one 9T-\rcam cell and the per-column peripherals in the figure, our array design has 256 9T-\rcam cells connected to a common column. These memory cells share 1 RAM SA, as well as 2 pre-charge and 2 search transistors. Furthermore, the array has 8 columns. Memory cells in the same row and different columns share 1 CAM SA. \rcam has two access modes. Below, we detail the operation of these two modes.%Furthermore, we discuss how each mode is leveraged in AES-based encryption/decryption. 

\subsection{RAM Mode}
\label{sec:RAM_mode_write_read}

Our 256$\times$8 \rcam array performs writes and reads in the RAM mode. The writes and reads with the 9T-\rcam are similar to the write and read operations with a 6T-SRAM (e.g., \cite{si19}). At the end of a write operation, the nodes $n_{1}$ and $n_{2}$ store opposite logic states. The transistors $M_{1}$, $M_{2}$, and $M_{3}$ are not used in the RAM mode.

%To write or read a row of the \rcam array, we first input a 8-bit memory address (i.e., the row index from 0x00 to 0xFF) to the row decoder. Based on this address, the circuits of the row decoder activate one specific row of the array by asserting its wordline ($WL$) voltage to $V_{DD}$.

%Once the $WL$ is activated, we apply a $V_{DD}$ voltage to the bitline ($BLWR$) to write the memory word. The opposite voltage  (i.e., $0V$) is applied to $\overline{BLWR}$. 
%At the end of the write operation, the nodes $n_{1}$ and $n_{2}$ store opposite logic states. For a read operation, $BLWR$ and $\overline{BLWR}$ need to be pre-charged to $V_{PRE}$ prior to $WL$ activation. In our design, we adopt $V_{PRE}=0.5V$. Pre-charging is performed by asserting $\overline{PRE}=0V$. After the pre-charging process and the $WL$ activation, the RAM sense amplifier is activated by setting $SENSE$ $RAM$ equal to $V_{DD}$. Then, the RAM sense amplifier circuit senses and amplifies the voltage drop at $BLWR$, in order to generate the output $OUT$ $RAM$ $SA$.

\subsection{CAM Mode}

The CAM mode enables a parallel search for any byte value in the range 0x00--0xFF among the 256 bytes stored in the 256$\times$8 \rcam array. The search drivers in Fig. \ref{fig:array}(a) distribute the byte pattern to the 8 columns of the array (through $SL$ and $\overline{SL}$). Before a search operation, a $0V$ voltage pulse is applied to $BLWR$ and $\overline{BLWR}$, to ensure that any residual voltage at the node $n_{3}$ is discharged (Fig. \ref{fig:array}(b)). To initiate a search, we assert $SENSE$ $CAM=V_{DD}$, which activates the search transistors connected to $SL$ and $\overline{SL}$, and precharges the matchline $ML$ to $V_{DD}$.

As mentioned in Sec. \ref{sec:RAM_mode_write_read}, after a memory cell is written, the nodes $n_{1}$ and $n_{2}$ store opposite logic states, which implies that either the transistor $M_{1}$ or the transistor $M_{2}$ are turned on during search (i.e., $M_{1}$ and $M_{2}$ are never turned on simultaneously). At this point, if the logic state stored in $n_{1}$ ($n_{2}$) is equal to the logic state applied to the ${SL}$ ($\overline{SL}$), the node $n_{3}$ remains discharged and the $M_{3}$ transistor remains off. In this scenario, the $ML$ is not discharged, producing a match at the output of the CAM SA ($OUT$ $CAM$ $SA=V_{DD}$).

Alternatively, if the state stored in $n_{1}$ ($n_{2}$) is the opposite of the state applied to the ${SL}$ ($\overline{SL}$), there is a path from  $SL$ ($\overline{SL}$) to $n_{3}$ (through either $M_{1}$ or $M_{2}$), which enables the node $n_{3}$ to charge up and the transistor $M_{3}$ to turn on. In this scenario, the $ML$ is discharged, producing a mismatch at the output of the CAM SA (i.e., $OUT$ $CAM$ $SA=0V$).

\subsection{Step Combination in \imcrypto}
\label{sec:merged_steps}

To maximally exploit IMC and reduce computation cost, we leverage a step combination technique in \imcrypto. Specifically, we store the values of $1 \times sbox(a_{i,j})$ in the 4 \rcam arrays, and the values of $2 \times sbox(a_{i,j})$, and $3 \times sbox(a_{i,j})$ in the 8 RAM arrays of 1 LUT fabric module (instead of simply storing $sbox(a_{i,j})$). These values are needed by the matrix multiplication operation in the MixColumns step, hence storing them directly in the \rcam arrays enables us to easily combine the SubByte and MixColumns steps of AES encryption. The InvSubBytes, AddRoundKey, and InvMixColumns steps of AES decryption are also executed in a combined fashion, with the modification of peripheral circuits (i.e., the customized encoder) of the \rcam array. 

\subsubsection{Encryption}Fig. \ref{fig:merged}(a) illustrates the execution of the SubBytes and MixColumns in a LUT fabric module, with the RAM arrays and \rcam working in the RAM mode. Bytes $a_{i,j}$ in the same column of the input state array (i.e., bytes that have the same $j$ index) are given as memory addresses to the row decoders of the 8 RAM arrays and 4 \rcam arrays in a transposed fashion. In the example of Fig. \ref{fig:merged}(a), we use $j=0$. A pre-computed byte in each array is read out by the RAM SA at the given address. %Each \rcam array stores 256 pre-computed bytes in the range 0x00--0xFF for the following functions: $1 \times sbox(a_{i,j})$, $2 \times sbox(a_{i,j})$, and $3 \times sbox(a_{i,j})$. Note that instead of the $sbox(a_{i,j})$ substitutions, we directly store the multiplied substituted bytes in the \rcam arrays, so the SubBytes and MixColumns steps are performed in one shot. 
Once the bytes are read out from the \rcam arrays, they are XORed by the XOR tree to produce the result of the matrix multiplication in the MixColumns step for the column $j$ in the output state array. %For instance, when $j=0$:

% \begin{math}

% \centering
% \resizebox{1\hsize}{!}{%
% b_{0,0}= 2\! \times\! sbox(a_{0,0})\! \oplus\! 3\! \times\! sbox(a_{1,0})\! \oplus\! 1\! \times\! sbox(a_{2,0})\! \oplus\! 1\! \times\! sbox(a_{3,0})     }
% \resizebox{1\hsize}{!}{%
% b_{1,0}= 1\! \times\! sbox(a_{0,0})\! \oplus\! 2\! \times\! sbox(a_{1,0})\! \oplus\! 3\! \times\! sbox(a_{2,0})\! \oplus\! 1\! \times\! sbox(a_{3,0})     }
% \resizebox{1\hsize}{!}{%
% b_{2,0}= 1\! \times\! sbox(a_{0,0})\! \oplus\! 1\! \times\! sbox(a_{1,0})\! \oplus\! 2\! \times\! sbox(a_{2,0})\! \oplus\! 3\! \times\! sbox(a_{3,0})     }
% \resizebox{1\hsize}{!}{%
% b_{3,0}= 3\! \times\! sbox(a_{0,0})\! \oplus\! 1\! \times\! sbox(a_{1,0})\! \oplus\! 1\! \times\! sbox(a_{2,0})\! \oplus\! 2\! \times\! sbox(a_{3,0})     }
% \vspace{1ex}
% \end{math}

\subsubsection{Decryption} To avoid unnecessary power and delay overheads, \imcrypto does not require \rcam arrays to be re-programmed when the accelerator switches its functionality from encryption to decryption. Therefore, during decryption, the 4 \rcam arrays of 1 LUT fabric module contain the same values of encryption (i.e., $1 \times sbox(a_{i,j})$). The RAM arrays in the LUT fabric module are not used in AES decryption and can be turned off. 

Fig. \ref{fig:merged}(b) depicts IMCRYPTO's approach for combining steps in AES decryption. \imcrypto performs InvSubBytes in the periphery of the \rcam array operating in CAM mode (i.e., with customized encoders). Customized encoders allow the execution of InvSubBytes to be combined with AddRoundKey and InvMixColumns to enable efficient AES decryption. We divide the execution of combined steps in AES decryption with \imcrypto into stages. \textbf{Stage 1} correspond to the operations of InvSubBytes, and leverages the invertible nature of the Sbox function to perform the SubBytes and InvSubBytes with distinct modes of a \rcam array. \textbf{Stage 2} produces the result of AddRoundKey. Finally, \textbf{stage 3} computes the multiplication of InvMixColumns. 

In the example of Fig. \ref{fig:merged}(b), bytes $a_{i,j}$ with the same $j$ index are given as search data to the search drivers of the 4 \rcam arrays in a transposed fashion (similar to encryption). Although we illustrate only the operations in the last \rcam arrays in our LUT fabric module (i.e., where the search data is $a_{3,0}$), identical operations occur simultaneously in the other columns of arrays ($a_{0,0}$,$a_{1,0}$, and $a_{2,0}$). In \textbf{stage 1} of our approach, we search for the byte $a_{3,0}$ in the \rcam arrays. The CAM SAs detect a single match for each array (i.e., while 1 row of the \rcam array produces a match, 255 rows produce mismatches). Since the \rcam array stores $1 \times sbox(a_{i,j})$, the row which returned a match is encoded into a byte as the equivalent result of the InvSubBytes step.

%%%%%%%%%%%%%%%%%% from Sec. 5
%pseudo code for sha-256
\begin{algorithm}[t]
\caption{SHA-256: Message Padding}
\label{alg:sha256_1}
    \textbf{Input:} Initial message $w$\\
    \textbf{Output:} Padded message $w$\\
\begin{algorithmic}[1]
   % \State $\sigma0(A):=($\textsc{a RightRotate 7}$)\oplus$
    %    \State $($\textsc{a RightRotate 18}$)\oplus$
     %   \State $($\textsc{a RightShift 3}$)$;
    %\State  $\sigma1(A):=($\textsc{a RightRotate 17}$)\oplus$
     %   \State $($\textsc{a RightRotate 19}$)\oplus$
      %  \State $($\textsc{a RightShift 10}$)$;
        
    \For{$i=16$, $i<64$, $i++$}
        \State $s0:=\sigma0(w[i-15])$;
        \State $s1:= \sigma1(w[i-2])$
        \State $w[i]:=(w[i-16]+s0+w[i-7]+s1$ 
    \EndFor
    
    \State \textbf{return} $w$
%\EndProcedure
\end{algorithmic}
\end{algorithm}

\begin{algorithm}[t]
\caption{SHA-256: Main Hashing Function}
\label{alg:sha256_2}
    \textbf{Input:} Padded message $w$, initial hashes $A-H$, constant $k$\\
    \textbf{Output:} Resulting Hashes $A$,$B$,$C$,$D$,$E$,$F$,$G$,$H$\\
    %\textsc{Hashing}($w$,$A$,$B$,$C$,$D$,$E$,$F$,$G$,$H$,$k$)
\begin{algorithmic}[1]
     %\State $\Sigma0(A):=($\textsc{a RightRotate 2}$)\oplus($\textsc{a RightRotate 13}$)\oplus %($\textsc{a RightRotate 22}$)$;
     %\State $\Sigma1(A):=($\textsc{a RightRotate6}$)\oplus($\textsc{a RightRotate 11}$)\oplus %($\textsc{a RightRotate 25}$)$;
     %\State  $Majority(A,B,C):=A\cdot B\oplus A\cdot C\oplus B\cdot C$;
     %\State  $Choice(A,B,C):=A\cdot B\oplus\overline{A}\cdot C$;
    \For{$i=16$, $i<64$, $i++$}
        \State $s0:=\Sigma0(A)$;
        
        \State $s1:=\Sigma1(E)$;
        
        \State $ch:=Choice(E,F,G)$;
        \State $maj:=Majority(A,B,C)$;
        \State $temp1:=H+s1+ch+k[i]+w[i]$;
        \State $temp2:=s0+maj$;
        \State $H:=G$;
        \State $G:=F$;
        \State $F:=E$;
        \State $E:=D+temp1$;
        \State $D:=C$;
        \State $C:=B$;
        \State $B:=A$;
        \State $A:=temp1+temp2$;
    \EndFor
    
    \State \textbf{return} $A$,$B$,$C$,$D$,$E$,$F$,$G$,$H$
%\EndProcedure
\end{algorithmic}
\end{algorithm}
%%%%%%%%%%%%%%%%%%%%%%%%

In \textbf{stage 2}, the customized encoder simply performs an XOR operation between each InvSubBytes result and $rk_{0,0}$ (i.e., a byte of the round key) in order to produce the result of the AddRoundKey step. In \textbf{stage 3}, the 8-bit result of the AddRoundKey step is multiplied by $9$, $11$, $13$, and $14$. This multiplication can be done with combinational logic inside the customized encoder that maps one byte value into another.

Similar to encryption, during decryption the 4 8-bit outputs of the customized encoder at stage 3 are XORed with the other outputs of the first row of \rcam arrays in the LUT fabric module (i.e., those where search data inputs were $a_{1,0}$, $a_{2,0}$, and $a_{3,0}$) with the XOR tree. The final result, which corresponds to the output of the InvSubBytes step, is forwarded to the next block in the AES decryption datapath.

\section{A Use Case of \imcrypto Beyond AES}
\label{sec:block_ciphers_example}
%%%%%%%%%%%%%%%%%%%%%%%%%%%%%%%%%%%%%%%%%%%%%%%%%%%%%%%%%%%%%%%%%%%

%%%%%%%% New content

 \begin{figure}[t]
  \centering
  \includegraphics[scale=0.22]{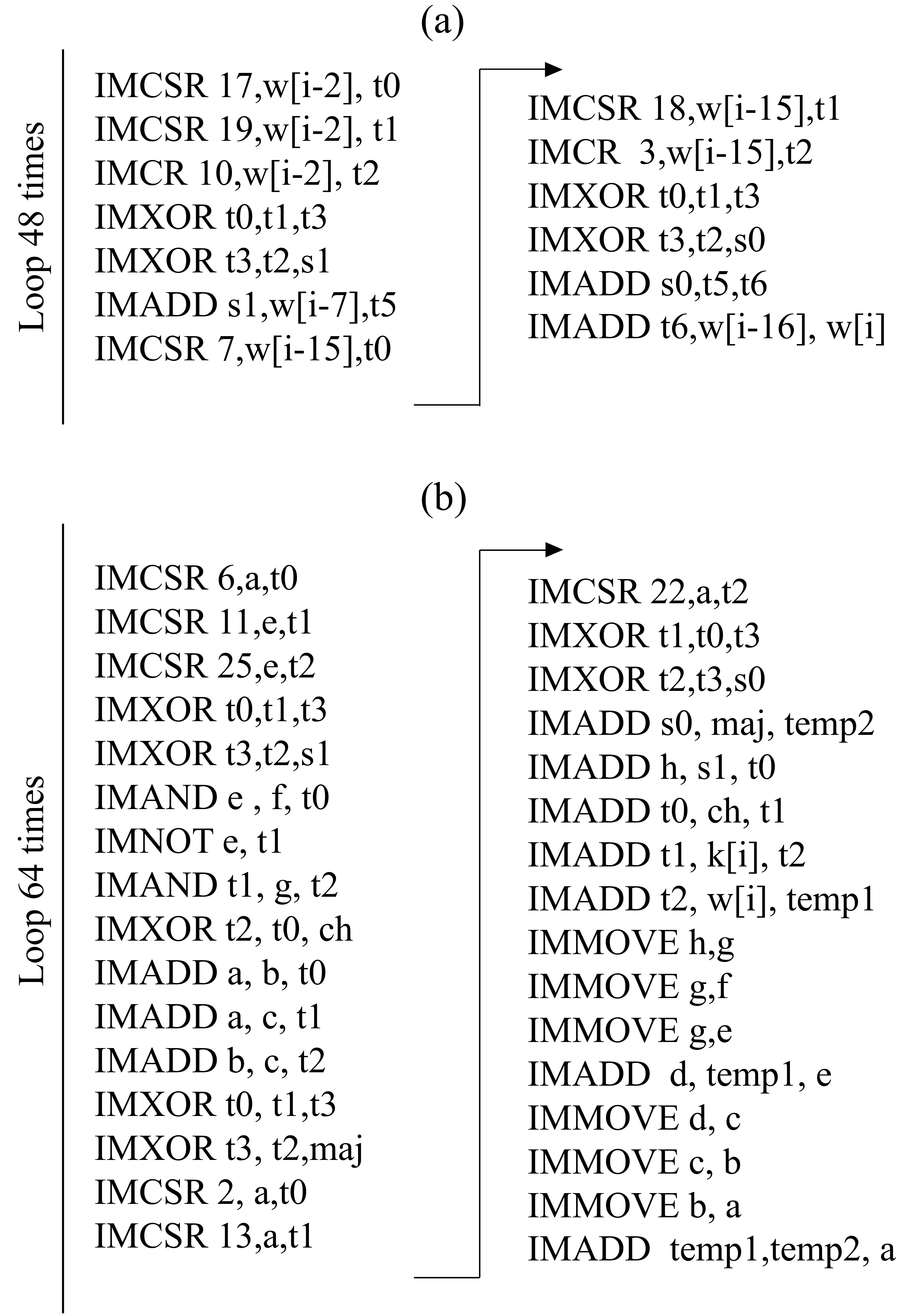}
  \vspace{-1ex}
  \caption{Translation of SHA-256 (a) message padding (Algorithm \ref{alg:sha256_1}) and (b) main hashing function (Algorithm \ref{alg:sha256_2}) into customized instructions.}
    %\vspace{-5ex}
    \label{fig:sha256_example}
\end{figure}

\bluHL{The distinct security needs by various applications, in addition to the development of new algorithms for handling emerging attacks make it highly desirable for crypto engines to be flexible. Furthermore, configurable crypto engines that can ``guarantee interoperability between countries and institutions", are highly sought-after intellectual property (IP) blocks \cite{bossuet13}. In \imcrypto, the use of a RISC-V co-processor allows for basic in-memory operations to be executed in an arbitrary order, which enables support for different modes of AES and other encryption/decryption algorithms beyond AES (e.g., hashing functions and other symmetric block ciphers). In this section, we provide an  example for implementing SHA-256 on our \imcrypto fabric.} 

\bluHL{SHA-256 is a member of SHA-2 --- a group of cryptographic hash functions designed by the National Security Agency (NSA). Widely known for its high security and speed, SHA-256 is used in the block chain industry and encryption communication protocols. The SHA-256 algorithm can be divided into two parts: \textit{message padding} (Algorithm \ref{alg:sha256_1}) and \textit{main hashing function} (Algorithm \ref{alg:sha256_2}). While the former extends a 16-bit word message to a 64-bit word message, the latter generates fixed-sized outputs (i.e., hashes). Message padding uses the $\sigma_{0}$ and $\sigma_{1}$ functions for padding (lines 2 and 3 in Algorithm \ref{alg:sha256_1}). $\sigma_{0}$ and $\sigma_{1}$ are implemented with bitwise ``right-rotate'' and ``right-shift" operations, as defined below:
\[ \sigma_{0}(A) = (A >>> 7) \oplus (A >>> 18) \oplus (A >>3) \]
\[ \sigma_{1}(A) = (A >>> 17) \oplus (A >>> 19) \oplus (A >>10) \]
\imcrypto uses the customized instructions IMCSL, IMCSR, IMSR, IMXOR and IMADD to perform message padding (Fig. \ref{fig:sha256_example}(a)). For instance, to implement rotations inside \imcrypto, the data are accessed through their memory addresses. Then, this data is read through the customized sense amplifier of the CEM and sent to hardware blocks that implement in-memory bit shifters. Fixed shift amounts are used with the bit shifters in order to slide the data bits to specific positions as required by the rotation operation.} 

\bluHL{The main hashing function (Algorithm \ref{alg:sha256_2}) is executed after message padding and produces the final output (i.e., a set of 8, fixed-sized hashes). The main hashing function in SHA-256 takes 64 rounds to complete its execution. It is comprised of two functions --- ``majority" and ``choice":
\[ Majority(A,B,C) = A\cdot B\oplus A\cdot C\oplus B\cdot C \]
\[ Choice(A,B,C):=A\cdot B\oplus\overline{A}\cdot C \]
Each of these functions take 3 inputs A, B, and C, and performs computation as follows: \textbf{(1)} The ``majority" function outputs `0'(`1') when half or more of its inputs are `0'(`1'). \textbf{(2)} The ``choice" function uses input A as a selector bit, i.e., when A=`1'(`0'), the function outputs the value of `B'(`C').  The main hashing function uses the $\Sigma_{0}$ and $\Sigma_{1}$ functions to hash the data, which can be implemented with bitwise ``right-rotate'' operations:
\[ \Sigma_{0}(A) =(A >>> 2)\oplus (A >>> 13) \oplus (A >>> 22)  \]
\[ \Sigma_{1}(A) =(A >>> 6)\oplus (A >>> 11) \oplus (A >>> 25)  \]
Besides these functions, 64 constants + 8 initial hashes (A -- H) are needed. \imcrypto pre-stores these constants and the initial hashes inside the CEM, and uses the customized instructions IMCSR, IMXOR, IMAND, IMOR, IMNOT, IMADD and IMMOVE (Table \ref{tab:instructions_type_r}) to perform the operations in the main hashing function.}

%%%%%%%%%%%%%%%%%%%%%%%%%%%%%%%%%%%%%%%%%%%%%%%%%%%%%%%%%%%%%%%%%%%
\section{Experimental Evaluation}
\label{sec:evaluation}
%%%%%%%%%%%%%%%%%%%%%%%%%%%%%%%%%%%%%%%%%%%%%%%%%%%%%%%%%%%%%%%%%%%

\begin{table}[t]
\caption{Accelerators Considered in the Evaluation}
\begin{tabular}{|c|c|c|c|c|}
\hline
\textbf{Design} & \textbf{Type} & \textbf{\begin{tabular}[c]{@{}c@{}}Technology\\ Node\end{tabular}} & \textbf{\begin{tabular}[c]{@{}c@{}}Supported\\ Algorithms\end{tabular}}                    & \textbf{Reference} \\ \hline
A               & ASIC          & CMOS 45 nm               & \begin{tabular}[c]{@{}c@{}}AES encryption/\\ decryption\end{tabular}                       & \cite{ueno20}                  \\ \hline
B               & ASIC          & CMOS 45 nm               & \begin{tabular}[c]{@{}c@{}}AES encryption+ \\ Most symmetric\\ciphers\end{tabular} & \cite{sayilar14}                  \\ \hline
C               & IMC           & CMOS 40 nm               & \begin{tabular}[c]{@{}c@{}}AES encryption,\\ Keccak, finite\\ field multiply\end{tabular}  & \cite{zhang18}                  \\ \hline
D               & IMC           & PCM 65 nm                & \begin{tabular}[c]{@{}c@{}}AES encryption/\\ decryption \end{tabular}                      & \cite{xie18}                  \\ \hline
E               & IMC           & CMOS 45 nm               & \begin{tabular}[c]{@{}c@{}}AES encryption/\\ decryption$\mathbf{{}^\star}$\end{tabular}                      & IMCRYPTO           \\ \hline
\end{tabular}
\label{tab:table_comparison}
\\
\begin{flushleft}
{\bf Notes:} $\mathbf{{}^\star}$\imcrypto can potentially support other symmetric ciphers by manipulating its custom instructions/reordering the steps executed in the IMC architecture. \\
\end{flushleft}
%\vspace{-4ex}
\end{table}

\begin{figure*}[!t]
  \centering
  \includegraphics[scale=0.15]{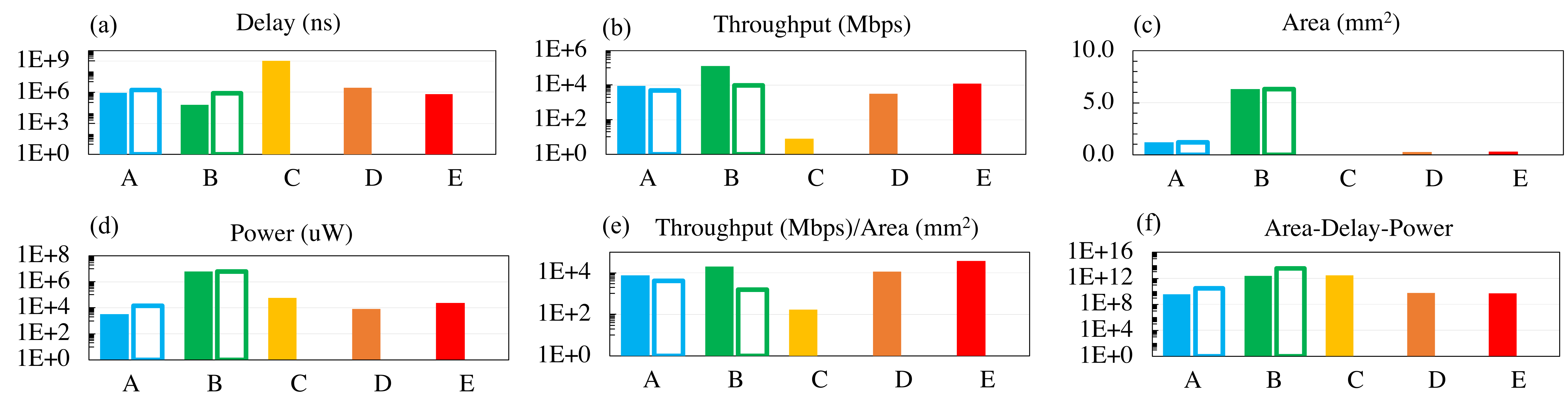}
  %\vspace{-1ex}
  \caption{(a) Delay, (b) throughput, (c) area, (d) power,  (e) throughput per area, and (f) area-delay-power product (ADPP) of AES-128 encryption. We compare the AES accelerators listed in Table \ref{tab:table_comparison}. Filled (hollow) bars do not (do) include the latency and energy overhead of data transfers to/from a memory unit for the ASIC-based accelerators (designs A and B).}
  \label{fig:results}
    %\vspace{-2ex}
\end{figure*}

 \begin{figure}[!t]
  \centering
  \includegraphics[scale=0.15]{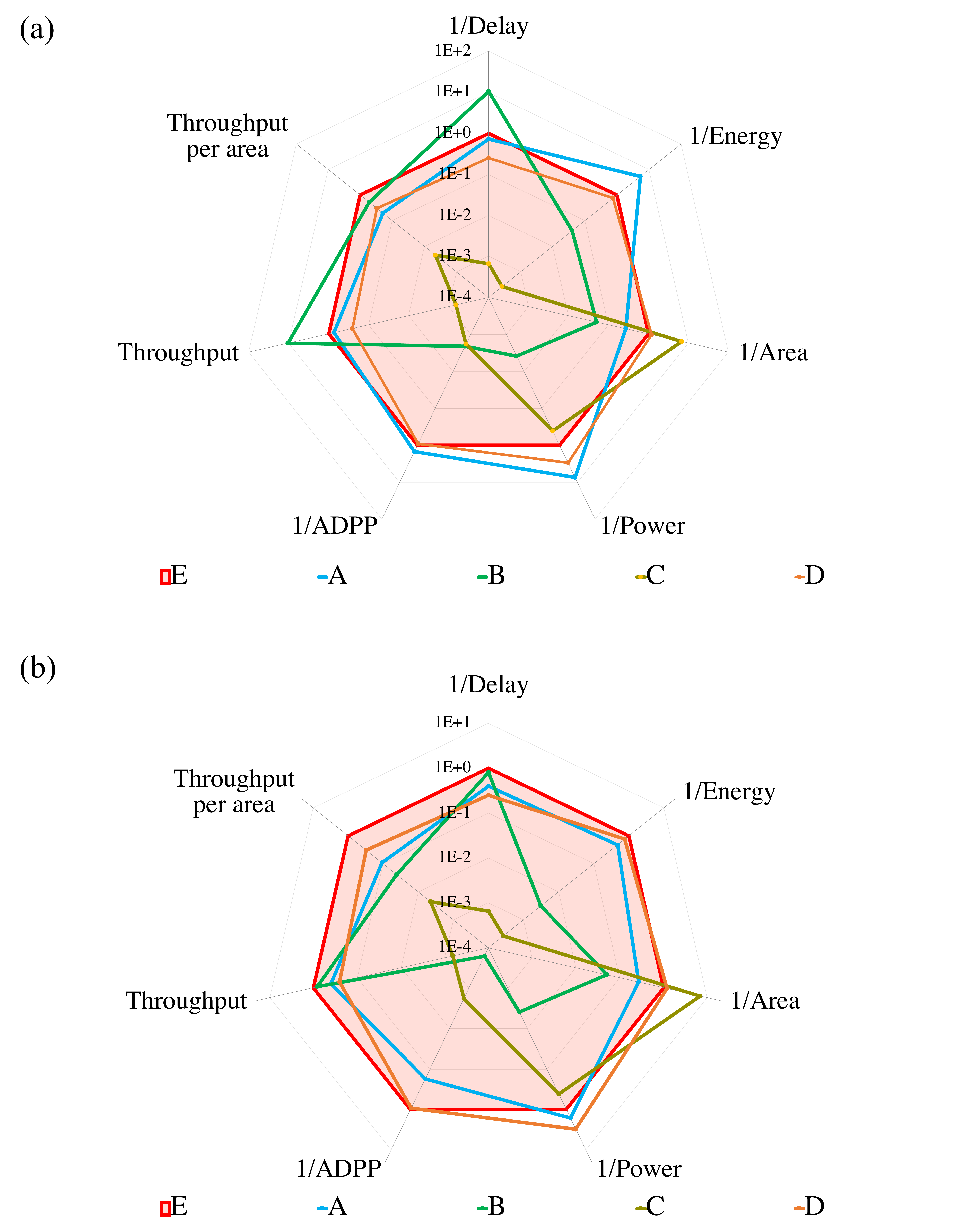}
  %\vspace{-1ex}
  \caption{Comparison between designs A-E for 7 different FoM (a) without and (b) with the overhead of data transfers.}
    %\vspace{-5ex}
    \label{fig:radar}
\end{figure}

 \begin{figure}[t]
  \centering
  \includegraphics[scale=0.23]{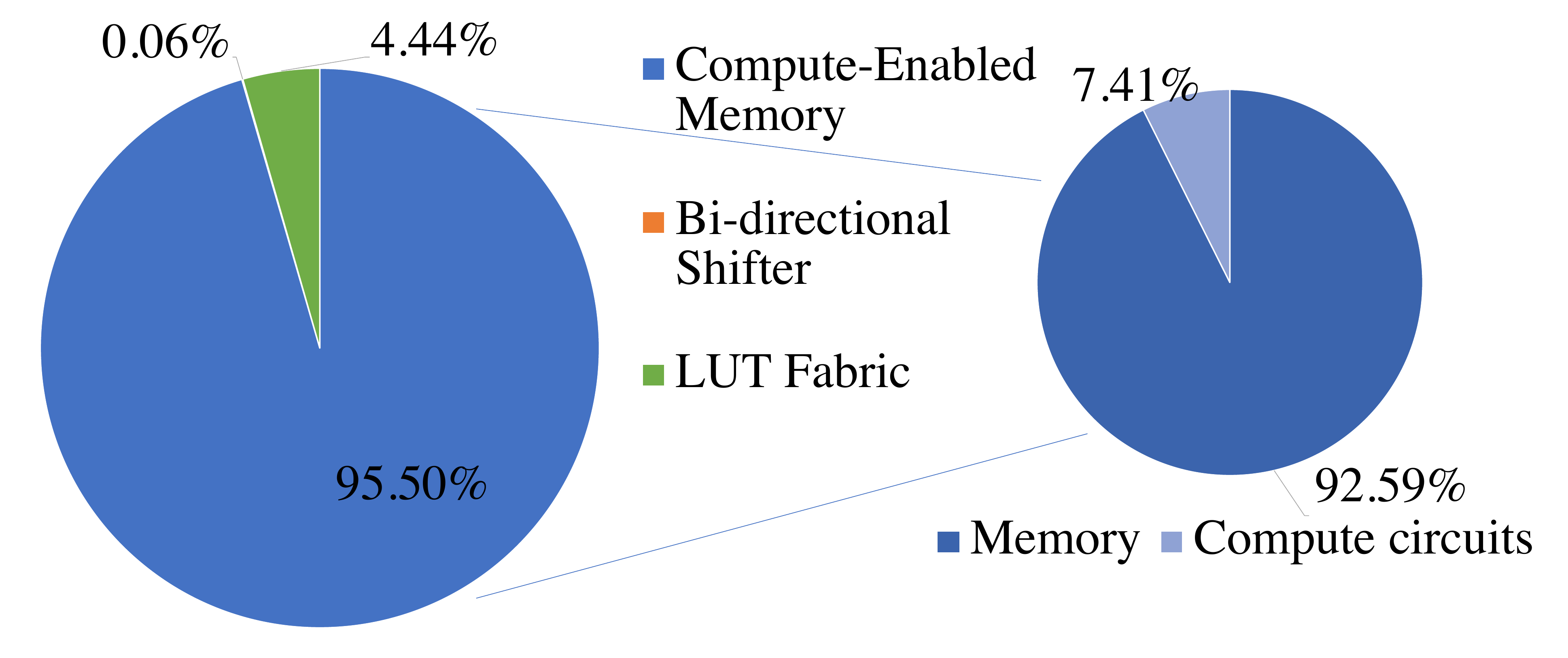}
  %\vspace{-1ex}
  \caption{Area breakdown of the \imcrypto fabric based on a 1 MB CEM. Memory components dominate the overall area of the proposed accelerator.}
    %\vspace{-1ex}
    \label{fig:area_breakdown}
\end{figure}

Here, we compare figures-of-merit (FoM) (i.e., delay, throughput, area, power, throughput per area, and ADPP) of \imcrypto with accelerators for AES encryption/decryption from previous works \cite{ueno20, sayilar14, zhang18}.

\subsection{Experimental Setup}

%Our approach that combines steps in AES encryption/decryption is validated with a in-house software implementation that employs combined tables for (Inv)SubBytes and (Inv)MixColumns steps of AES. 

In our evaluation (and comparison), \textbf{we implement AES-128 encryption (ECB mode).} The implementation of the \imcrypto instructions in the RISC-V based controller and their sequencing for AES-128 ECB mode is done in Verilog and evaluated through simulations. At the circuit level, we use the DESTINY simulator \cite{destiny}\footnote{DESTINY is an open-source tool to simulate 2D and 3D memory arrays, which utilizes the 2D modeling framework of NVSim~\cite{dong12_nvsim} for SRAM and non-volatile memories based on emerging technologies.} to measure area, latency and energy for a 1MB compute-enabled SRAM, with a 2-mat organization and a subarray size of 128$\times$128. We have modified DESTINY to support the customized sense amplifiers and local write buffers of \cite{aga17}. The compute-enabled SRAM is based on a 45nm CMOS predictive technology model (PTM) \cite{cao2011predictive}. We choose the 45nm technology node as most designs in our comparison use this node. To estimate the area overhead introduced by compute circuits in the SRAM, we subtract the area of a baseline SRAM of 1 MB based on same technology node from the area of the compute-enabled RAM. We use DESTINY to simulate the baseline SRAM, which is a regular SRAM memory without customized peripherals.

The area, latency, and energy of the bi-directional shifter block are measured through synthesis using the Cadence Encounter RTL Compiler v14.10, with the NanGate 45nm open-cell library \cite{knudsen2008nangate}. Finally, the 256$\times$8 \rcam arrays in \imcrypto's LUT fabric (block 4 of Fig. \ref{fig:architecture}(a)) are simulated using HSPICE version O-2018.09-1 with the PTM for 45nm CMOS\cite{cao2011predictive} and $V_{DD}=1V$. We measure the latency and energy for the different types of memory accesses, i.e., reads/writes in RAM mode and searches in CAM mode. In order to estimate the area of the 256$\times$8 \rcam array, we employ the OpenRAM memory compiler \cite{guthaus2016openram}, in which we specify the 6T-SRAM and 9T-\rcam cells as well as the peripherals of our design. The area of the \imcrypto fabric (which we compare with the area of other accelerators in Sec. \ref{sec:fom_analysis}) includes the area overhead of compute-enabled SRAM, the area of the bi-directional shifter, and the area of all circuits in the LUT fabric.

\subsection{Quantitative Analysis with Different FoM}
\label{sec:fom_analysis}

Table \ref{tab:table_comparison} summarizes previous work on accelerators for AES, along with our proposed design \imcrypto (design E). Designs A and B are ASIC accelerators for AES based on a 45nm CMOS technology node. Designs C, D and E are based on the IMC concept and consist of CEM arrays integrated with customized circuitry that compute at the memory periphery. While designs C and E are based on 40nm and 45nm CMOS technologies respectively, design D is based on a 65nm PCM technology. Designs B and C support other algorithms/functions beside AES encryption. Support for AES decryption is not discussed in \cite{sayilar14,zhang18}. AES decryption is possible with designs A, D and E.% Designs B---E implement ECB mode, while design A implements CTR mode.

Fig. \ref{fig:results}(a-f) and Fig. \ref{fig:radar}(a-b) present and compare different FoM for designs A-E for AES-128 encryption. While Fig. \ref{fig:results}(a-f) presents the raw data for all FoM in separate bargraphs, Fig. \ref{fig:radar}(a-b) enables us to collectively compare the different designs in terms of all FoM evaluated. Two sets of data are considered: The filled bars in Fig. \ref{fig:results}(a-f) (lines in Fig. \ref{fig:radar}(a)) \textit{do not} include the latency and energy of data transfers to/from a memory unit for the two ASIC-based accelerators (designs A and B). Note that designs C---E are based on the IMC paradigm, which allow computation to be performed inside (or near) the memory without the need for data transfers. The second set of data (hollow bars in Fig. \ref{fig:results}(a-f)/ lines in Fig. \ref{fig:radar}(b)) adds 744.49$\mu$s latency (30.39$\mu$J energy) from memory transfers to the latency (energy) of designs A and B. This additional latency/energy is spent on transferring 1MB of plaintext from a 1MB SRAM to each ASIC accelerator, and writing back the resulting 1MB of ciphertext to the same 1MB SRAM. DESTINY \cite{destiny} is used to estimate the latency and energy overheads of such data transfers.

We first consider delay and throughput when data transfers are \textit{not} accounted for in designs A and B (i.e., the ASIC-based accelerators). In this scenario, design B has the shortest delay to encrypt 1MB of data and the highest throughput among all designs evaluated (Fig. \ref{fig:results}(a) and Fig. \ref{fig:results}(b)). \imcrypto has the highest throughput per area (Fig. \ref{fig:results}(e)) among all designs in our study. Compared to designs A-D, design E enables minimum improvements in throughput per area of $\sim$5.0$\times$, $\sim$1.9$\times$, $\sim$223.1$\times$, and $\sim$3.3$\times$, respectively. 

When we include the overhead of data transfers for the ASIC-type accelerators in our analysis, design E has the shortest delay and highest throughput among all designs. Namely, when compared to design B, design E is 1.2$\times$ better with respect to both delay and throughput. Importantly, fast and high throughput AES-128 encryption in \imcrypto (design E) comes with an improvement of 19.7$\times$ in terms of area (Fig. \ref{fig:results}(c)) and 257.7$\times$ in terms of power (Fig. \ref{fig:results}(d)) compared to design B. Finally, when we account for data transfers, design E outperforms design B by a factor of 5.7$\times$ in terms of ADPP (Fig. \ref{fig:results}(f)).

Note that in terms of the ADPP metric, design A would be the best-in-class design when data transfers for ASIC-based designs \textit{are excluded} from the evaluation. In the setup where data transfers are included in the total delay and energy results, the second-best design in terms of ADPP is design D, an architecture also based on IMC. Unlike design E (which is based on CMOS), design D employs PCM and bank-level parallelism in its CEM. Design E slightly improves the ADPP by a factor of 1.2$\times$ compared to design D, which is associated with the two designs occupying roughly the same area, with design E having a 4.0$\times$ shorter delay and a higher power consumption (0.3$\times$). Some of the benefits of design D are drived from the use of PCM. In fact, \imcrypto can also benefit from alternative memory technologies which we discuss next.

\subsection{FoM Projection based on Emerging Technologies}

\begin{figure}[t]
  \centering
  \includegraphics[scale=0.225]{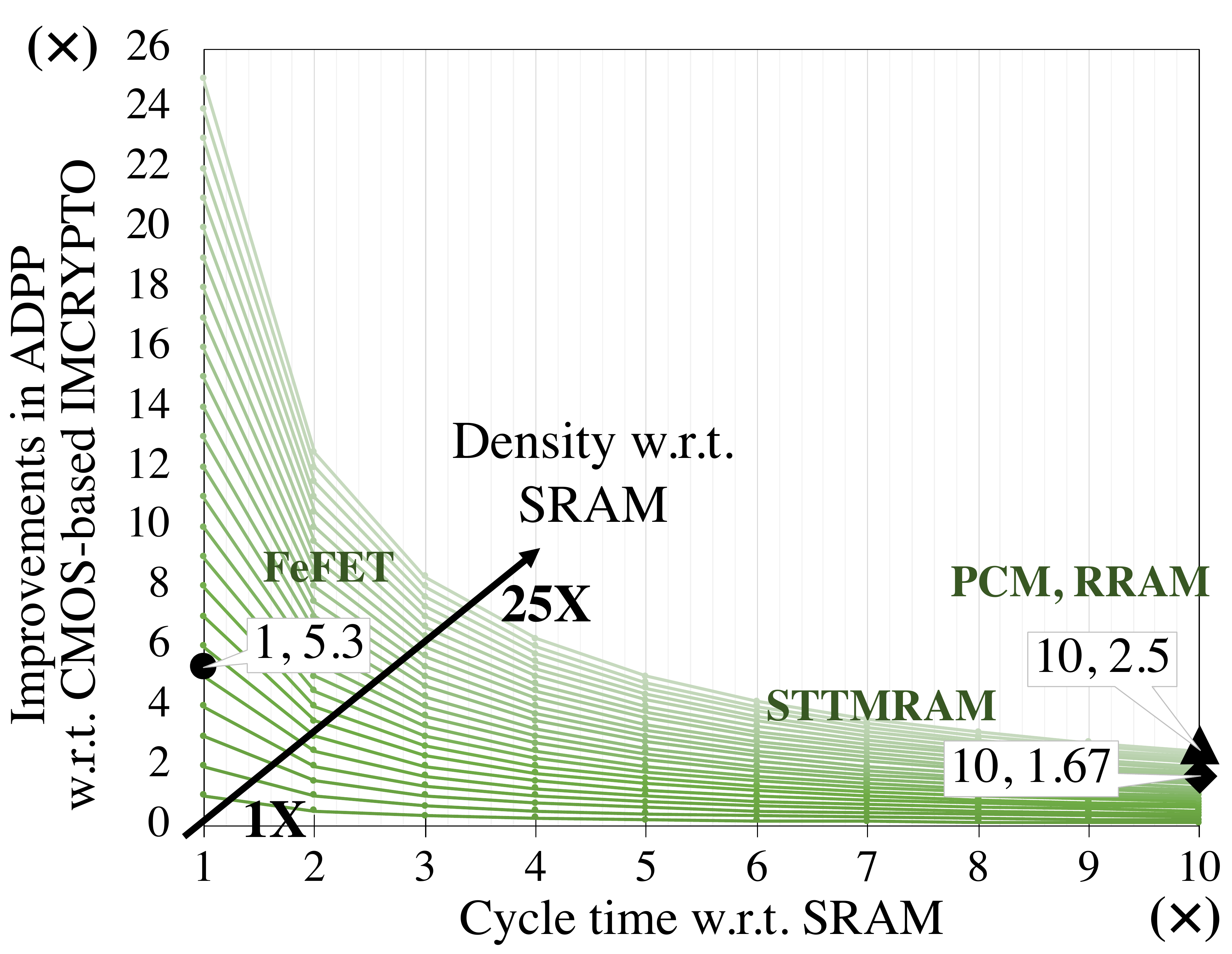}
  %\vspace{-1ex}
  \caption{Projected improvements in ADPP with respect to CMOS-based \imcrypto for memories based on emerging technologies. STT-MRAM, PCM and RRAM density and cycle time data points are from \cite{yu2016emerging}. FeFET density and cycle time data points are from \cite{reis19_1fefet}. The pair of numbers for each technology indicate the cycle time with respect to SRAM and improvements in ADPP with respect to CMOS-based \imcrypto.}
  \label{fig:emerging}
    \vspace{-2ex}
\end{figure}

Per Fig. \ref{fig:area_breakdown}, the CEM and the LUT fabric modules of \imcrypto are responsible for $>$99.9$\%$ of the total area of our proposed AES encryption/decryption IMC accelerator. As discussed in Sec. \ref{sec:cim_design}, these two blocks of \imcrypto are built with CMOS 6T-SRAM and 9T-\rcam cells. Here, we investigate whether the use of memories based on emerging technologies such as PCM, RRAM, STT-MRAM, and FeFETs to replace CMOS-based memories could result in additional benefits to our proposed \imcrypto approach.

Per \cite{yu2016emerging,sun2018memory,reis19_1fefet}, emerging technologies could potentially improve the density of memories by factors of 1$\times$--25$\times$. Furthermore, the read power of memories based on most emerging technologies may be comparable to that of CMOS SRAM, with static power savings due to non-volatility \cite{sun2018memory}. Despite these advantages, one of the downsides of emerging technologies is their long read and write latencies (up to 10$\times$ of CMOS SRAM), which may increase clock cycle time and reduce the frequency of operation of IMC accelerators such as \imcrypto. To analyze the impact of emerging technologies on the ADPP of \imcrypto, we study the trade-off between longer cycle times (thus lower operating frequencies) and density improvements enabled by emerging technologies with respect to CMOS SRAM (Fig. \ref{fig:emerging}). 

In our projection, we consider the read power of emerging technologies to be on par with that of CMOS SRAM \cite{sun2018memory}. The green curves in the plot of Fig. \ref{fig:emerging} represent different memory densities (1$\times$--25$\times$ compared to SRAM). Given cycle times that are up to 10$\times$ longer than those of conventional SRAM (x-axis), improvements in ADPP with respect to CMOS-based \imcrypto (y-axis) are captured. From the data points depicted in Fig. \ref{fig:emerging} (from \cite{yu2016emerging, reis19_1fefet}), we project that current state of the art memory technologies improve the ADPP of IMCRYPTO by 1.7$\times$--5.3$\times$. For instance, by implementing a PCM-based \imcrypto, we project 2.5$\times$ benefits in ADPP relative to a CMOS-based \imcrypto (design E in Table \ref{tab:table_comparison}). As CMOS-based \imcrypto's ADPP slightly outperforms PCM-based design D by a factor of 1.2$\times$, when we compare the two designs implemented with same technology, we expect design E to outperform design D in terms of ADPP by a factor of 3.0$\times$.

Despite the projections, we acknowledge that considerable challenges from the system/architecture perspective exist in employing emerging technologies for the design of memories and IMC-based accelerators \cite{mutlu13}. For instance, technologies may be at different stages of maturity and not be as easily scalable to smaller nodes as is CMOS technology. Nevertheless, as memory technologies continue to evolve, opportunities emerge with the study of new materials, the possibility of stacked memory, etc. The impact of these advancements on the performance of IMC for security applications is worthwhile to pursue in further investigations. 

\section{Conclusion and Future Work}
\label{sec:conclusion}
%%%%%%%%%%%%%%%%%%%%%%%%%%%%%%%%%%%%%%%%%%%%%%%%%%%%%%%%%%%%%%%%%%%

We propose \imcrypto, an IMC fabric for accelerating AES encryption and decryption that enables high throughput while avoiding large area/power overheads. \imcrypto employs a novel approach that combines multiple RAM and \rcam arrays with customized encoders in the design of a modular LUT fabric. With the proposed LUT fabric, the (Inv)SubBytes and (Inv)MixColumns steps from AES encryption (decryption) can be combined and performed in one shot, in a highly parallel fashion. Furthermore, integration with RISC-V allows \imcrypto to have the flexibility of doing different modes of operation and opens the door for the implementation of other ciphers that require operations that are similar those of AES. Our future work will evaluate the execution of different AES modes and key lengths with \imcrypto, and will extend our customized instructions to enable the execution of other symmetric-key ciphers. \bluHL{Finally, to reduce the number of customized RISC-V instructions used in \imcrypto, we will exploit different levels of control granularity by replacing some of subset of customized instructions with ``macro instructions" and provide special hardware control units for these macro instructions.  }

%and exploit the \imcrypto's performance and security level of doing difference symmetric-key algorithms for cryptography.

%\vspace*{-1ex}

 \section*{Acknowledgments}
 This work was supported in part by ASCENT, one of six centers in JUMP, a Semiconductor Research Corporation (SRC) program sponsored by DARPA.

%\pagebreak
\IEEEtriggeratref{12}
\bibliographystyle{./IEEEtran.bst}
%\bibliography{references}
\bibliography{references}

\end{document}